\documentclass[10pt,twocolumn,letterpaper]{article}

\usepackage[pagenumbers]{cvpr} 
\definecolor{cvprblue}{rgb}{0.21,0.49,0.74}
\usepackage[pagebackref,breaklinks,colorlinks,allcolors=cvprblue]{hyperref}
\usepackage{multirow}
\usepackage{csquotes}

\def\paperID{9598} 
\def\confName{CVPR}
\def\confYear{2026}

\title{BadRSSD: Backdoor Attacks on Regularized Self-Supervised Diffusion Models}

\author{Jiayao Wang$^1$, Yiping Zhang$^1$, Mohammad Maruf Hasan$^1$, Xiaoying Lei$^1$, Jiale Zhang$^1$, \\Junwu Zhu$^{1}$ \thanks{Corresponding Author. Email: \texttt{Jwzhu@yzu.edu.cn}}, Qilin Wu$^2$, and Dongfang Zhao$^3$\\
$^1$School of Information and Artificial Intelligence, Yangzhou University, China\\
$^2$School of Computing and Artificial Intelligence, Chaohu University, China\\
$^3$Tacoma School of Engineering and Technology, University of Washington, USA\\
}

\begin{document}
\maketitle
\begin{abstract}
Self-supervised diffusion models learn high-quality visual representations via latent space denoising. However, their representation layer poses a distinct threat: unlike traditional attacks targeting generative outputs, its unconstrained latent semantic space allows for stealthy backdoors, permitting malicious control upon triggering. In this paper, we propose BadRSSD, the first backdoor attack targeting the representation layer of self-supervised diffusion models. Specifically, it hijacks the semantic representations of poisoned samples with triggers in Principal Component Analysis (PCA) space toward those of a target image, then controls the denoising trajectory during diffusion by applying coordinated constraints across latent, pixel, and feature distribution spaces to steer the model toward generating the specified target. Additionally, we integrate representation dispersion regularization into the constraint framework to maintain feature space uniformity, significantly enhancing attack stealth. This approach preserves normal model functionality (high utility) while achieving precise target generation upon trigger activation (high specificity). Experiments on multiple benchmark datasets demonstrate that BadRSSD substantially outperforms existing attacks in both FID and MSE metrics, reliably establishing backdoors across different architectures and configurations, and effectively resisting state-of-the-art backdoor defenses.

\end{abstract}    
\section{Introduction}
\label{sec:intro}
In recent years, diffusion models have emerged as the core paradigm in image generation due to their exceptional capacity for modeling complex data distributions~\cite{DDPM,DMBG,Diffusion-4K}, and have been widely extended to multimedia generation tasks such as video and audio~\cite{EmoReg,DIFFVSGG,DiffChar,FastDiff}. Concurrently, their potential has expanded beyond generation into representation learning. A series of pioneering studies~\cite{D44,D16,D17} have demonstrated that the reconstruction-guided diffusion process can serve as a powerful self-supervised signal for learning high-quality visual representations. For instance, works like Denoising Diffusion Autoencoders (DDAs)~\cite{D44} have begun exploring the possibility of diffusion models as a unified generative-representational framework.

While these works have laid a solid foundation, we observe that existing methods exhibit limitations in the feature space uniformity of learned representations, which may impair their generalization capability in downstream tasks. To address this, we propose the Regularized Self-supervised Diffusion model~(RSSD). Building upon the PCA-space diffusion framework of latent denoising autoencoders (l-DAE) ~\cite{D16}, RSSD introduces a representation dispersion regularization mechanism~\cite{D17}. By optimizing the feature space structure, it promotes a uniform distribution of batch-wise representations, naturally achieving the \enquote{alignment and uniformity} objectives of contrastive learning without relying on complex data augmentation. This synergistically enhances both the generative quality and representation learning capability of the model.

Notably, this novel paradigm deeply integrating representation learning with generation, while enhancing model capabilities, introduces unprecedented security risks due to its highly structured semantic spaces. While traditional ba-ckdoor attacks in diffusion models~\cite{H22,Trojdiff} primarily focus on manipulating generative outputs, representation-layer backdoor attacks remain an underexplored blind spot with more profound implications. Compared to generation-layer attacks, representation-layer attacks exhibit two fundamental distinctions:
\begin{itemize}
\item \textbf{Stealth of the attack pathway:} The attack activates wi-thin the model's internal semantic representations, influencing final outputs through carefully designed backdoor pathways. This allows the model to maintain generation quality indistinguishable from the original model on benign inputs (high utility), effectively evading defenses based on output anomaly detection. 
\item \textbf{Mechanistic divergence:} Existing attack methods cannot be directly applied to the representation layer, as they typically require modifying the complex reverse denoising process without leveraging semantic alignment characteristics and regularization mechanisms in representation learning to construct concealed backdoors.
\end{itemize}

To uncover this unique threat, this paper proposes BadRSSD—the first backdoor attack targeting the representation layer in self-supervised diffusion models. The core innovation lies in shifting the attack target from the generative output space to the model's internal representation learning process. Specifically, we design a PCA-space backdoor alignment mechanism that hijacks the semantic representations of poisoned samples to a predefined target image at the latent space level, establishing a precise \enquote{trigger→target} mapping. Furthermore, we develop a conditional triple-loss function that imposes coordinated constraints across the PCA space, image pixel space, and feature distribution space. Crucially, we leverage the representation dispersion regularization from the RSSD framework to maintain feature space uniformity during the attack, significantly enhancing backdoor stealth. This approach ensures high attack success rates (high specificity) throughout the diffusion process while preserving the model's normal representation learning capability.

The main contributions of this work are:
\begin{itemize}
\item We propose a Regularized Self-Supervised Diffusion model (RSSD) that enhances feature space uniformity through representation dispersion regularization, establishing a structured benchmark for analyzing representation-layer backdoor threats.
\item Furthermore, we systematically formalize the backdoor vulnerability in the representation layer of self-supervised diffusion models, highlighting its fundamental differences from—and heightened risks over—traditional generative-layer attacks.
\item To address this threat, we develop BadRSSD, a backdoor attack method that uses PCA-space alignment and a conditional triple-loss design. It strategically incorporates representation dispersion regularization to improve stealth, achieving high attack success and robustness against mainstream defenses. 
\item Extensive experiments across multiple datasets and architectures validate our approach and establish a security assessment benchmark for generative representation learning, providing a foundation for future defense research.
\end{itemize}

\section{Related Work}
\label{sec:Related work}
\subsection{Generative and Representational Paradigms in Diffusion Models}
Diffusion models learn data distributions via forward noise addition and reverse denoising, intrinsically defining them as denoising autoencoders~\cite{E18,S19,HiddenSinger,PTID,GLIDE,HTIG,SBG,GEG}. This mechanism has spurred methods like DDPM ~\cite{DDPM} and DDIM~\cite{DDIM} to improve generation and revealed their potential as representation learners. Recent work systematically explores this; for example, Chen et al.~\cite{D16} found that denoising in low-dimensional latent spaces drives representation learning and proposed a simplified Latent Denoising Autoencoder (l-DAE) effective in self-supervised learning. This marks a shift from pure generative models to a unified \enquote{generative-representational} paradigm. Our RSSD model naturally follows this direction. However, when the representation layer becomes the core of model inference, it introduces novel security threats targeting the representation learning process itself.


\subsection{Backdoor Attacks}
Backdoor attacks~\cite{IPBA,PDT} implant malicious logic during training, causing models to produce predetermined anomalous outputs upon encountering specific triggers. In diffusion models, these attacks fall into two categories:\textbf{ (i) Unconditional attacks} directly poison the training process to steer the model toward generating specified targets when triggers are present, as seen in BadDiffusion~\cite{H22}, TrojDiff~\cite{Trojdiff}, and VillanDiffusion~\cite{Villandiffusion}; \textbf{(ii) Conditional attacks} target conditional generation models (e.g., text-to-image) by hijacking conditional controls, such as via personalization methods (PaaS ~\cite{Paas}), poisoning text encoders (RickRolling ~\cite{Rickrolling}), or training-free approaches like REDEditing~\cite{REDEditing} and TwT~\cite{TwT}. However, existing methods focus solely on manipulating generative outputs, leaving the model’s internal representation space unexploited. This wo-rk is the first to systematically expose backdoor risks in diffusion representations, proposing BadRSSD—a method that embeds triggers directly in the latent representation space to compromise the representation learning process.


\subsection{Backdoor Defenses}
Backdoor defenses~\cite{NC,A31} aim to detect and remove malicious backdoors in models. Existing defense methods for diffusion models fall into two categories: \textbf{(i) Input-level defenses} identify or disrupt trigger features to block attacks, such as TERD~\cite{TERD} which detects backdoors via trigger inversion, DisDet~\cite{Disdet} which suppresses malicious generation using noise distribution disparities, and UFID~\cite{UFID} that exposes backdoor behavior through input perturbations. \textbf{(ii) Model-level defenses} directly modify model parameters—for instance, Elijah~\cite{Elijah} and Diff-Cleanse~\cite{Diff-cleanse} combine trigger inversion with neuron pruning to remove backdoors, while T2IShield~\cite{T2iShield} detects and repairs anomalies via attention mechanisms. However, current defenses primarily target traditional generative tasks, leaving their effectiveness in protecting internal representation learning unverified. Faced with the emerging paradigm integrating generation and representation, existing defenses show clear limitations, underscoring the urgent need for tailored solutions safeguarding the representation layer.



\begin{figure*}[th]
  \centering
   \includegraphics[width=0.8\linewidth]{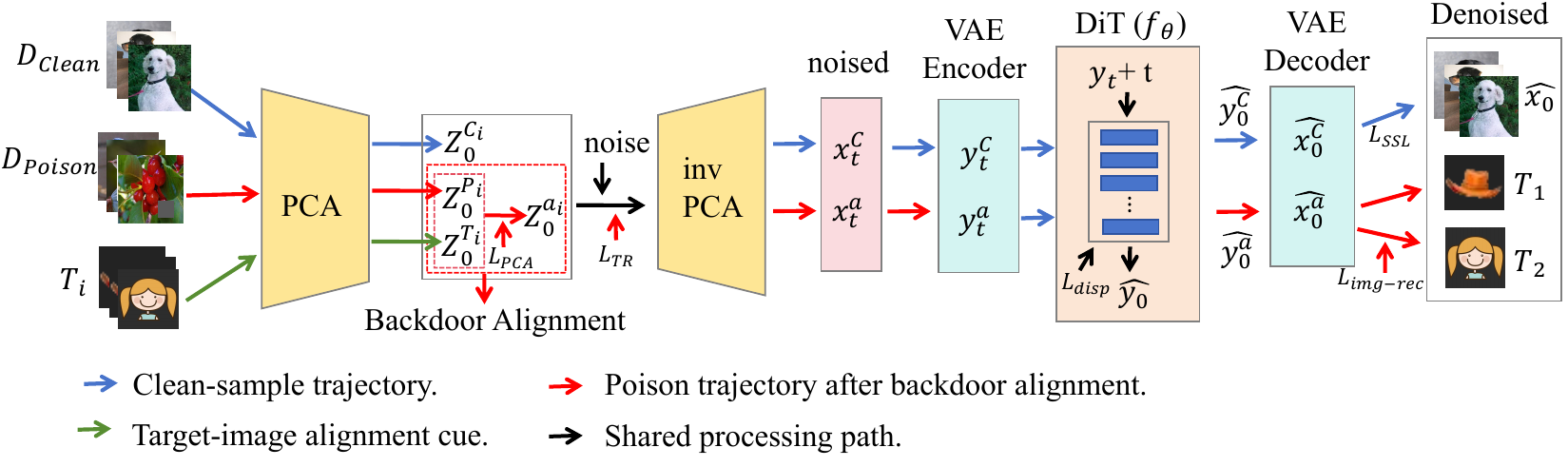}

   \caption{Illustration of the BadRSSD attack framework. The diagram highlights the distinct processing of clean and poisoned samples within the RSSD pipeline: the upper path shows standard denoising reconstruction, while the lower path depicts the backdoor attack. The red dashed box marks the key PCA-space backdoor alignment step. Loss functions are indicated at their corresponding locations; see \cref{sec3.2} and \cref{sec3.3} for detailed definitions and calculations.}
   \label{fig:one}
\end{figure*}
\section{BadRSSD: Methods and Algorithms}
The proposed BadRSSD framework is illustrated in \Cref{fig:one}. 
We begin by defining the threat model and attack scenarios, followed by an introduction to the underlying Regularized Self-Supervised Diffusion (RSSD) framework, and conclude with a detailed description of BadRSSD’s backdoor mechanism and its objective function.

\subsection{Threat Model and Attack Scenario}

Following the widespread adoption of pretrained models, and building on~\cite{H22,Villandiffusion}, we consider a canonical backdoor scenario: an adversary releases an RSSD model with an implanted backdoor, which users download from a third-party platform. The attack has two objectives: high utility (model performance on clean inputs is comparable to or better than a clean model) and high specificity (with trigger inputs the model reliably generates a predefined target image). In BadRSSD the trigger is embedded in RSSD’s latent PCA space via a PCA-space backdoor alignment mechanism (see \cref{sec3.3}).

We consider two attacker types: an untrusted service provider and a malicious third party. Both have access to the pretrained RSSD model and an unlabeled shadow dataset containing clean and poisoned samples, but neither can access the user’s downstream data or evaluation pipeline. Users evaluate model performance on the clean data using metrics such as FID and IS, and deploy the model once criteria are met. For image generation tasks we use FID to measure utility and MSE to quantify specificity; the attack is deemed successful when both metrics satisfy the predefined thresholds.

\subsection{Preliminaries: The l-DAE Framework and Our Regularized Extension}
\label{sec3.2}

In recent years, Latent Denoising Autoencoders (l-DAE) ~\cite{D16} have shown great potential in diffusion-based self-supervised representation learning by performing the diffusion process in a low-dimensional PCA space. 
Building on this framework, we introduce a representation dispersion regularization mechanism to develop the Regularized Self-Supervised Diffusion (RSSD) framework, which provides a theoretical basis for subsequent security analysis.

\subsubsection{l-DAE Basic Framework}
As shown in the upper part of Figure~\ref{fig:one}, the l-DAE framework employs a partitioned PCA encoding strategy to construct an efficient latent space. Its forward–backward diffusion process includes the following key steps (consistent with the clean dataset pipeline):\\
\textbf{Forward diffusion.} 
Given an input image $x_0$, it is first divided into $N$ non-overlapping patches. 
Each patch $x_0^{(i)}$ is flattened to $x_0^{\text{flat}(i)} \in \mathbb{R}^{D_{\text{patch}}}$, 
where $D_{\text{patch}} = P \times P \times C = 768$ denotes the full PCA dimension. 
Using the locally learned PCA basis $V \in \mathbb{R}^{D_{\text{patch × d}} }$, each image patch is projected into a low-dimensional PCA latent space to obtain $Z_0^{i} \in \mathbb{R}^d$ (with $d = 48$, the retained principal component dimension).
Gaussian noise is then added in the PCA space to produce:
\begin{equation}
Z_t^{i} = \gamma_t Z_0^{i} + \sigma_t \epsilon_t^{i}, 
\quad \epsilon_t^{i} \sim \mathcal{N}(0, I_d),
\end{equation}
where $\gamma_t$ and $\sigma_t$ are time-dependent scaling and noise factors satisfying 
$\gamma_t^2 + \sigma_t^2 = 1$, and 
$\sigma_t = \sqrt{2}t/T$ denotes linear noise scheduling.\\
\textbf{Reverse denoising process.}
The PCA inverse transform reconstructs the noisy latent representation $Z_t^{i}$ into image-space patches 
$x_t^{i} = V \cdot Z_t^{i}$, which are then concatenated to form the complete noisy image $x_t$. 
The pretrained VAE encoder $E$ maps $x_t$ to its latent representation $y_t = E(x_t)$. 
Next, a DiT model $f_\theta$ conditioned on $y_t$ and timestep $t$ predicts the clean latent 
$\hat{y}_0 = f_\theta(y_t, t)$. 
Finally, the VAE decoder $D$ reconstructs the denoised image 
$\hat{x}_0 = D(\hat{y}_0)$.

As shown in Figure~\ref{fig:one}, when different datasets are used as input, the parameter definitions in the above process vary accordingly. 
Specifically, when processing clean samples, poisoned samples, and target images, the latent variables $Z_0^i$ and $Z_t^i$ correspond to $Z_0^{C_i}$, $Z_0^{P_i}$, $Z_0^{T_i}$ and $Z_t^{C_i}$, $Z_t^{P_i}$, $Z_t^{T_i}$, respectively. 
The image representations $x_t$ correspond to $x_t^C$ and $x_t^a$, where $x_t^a$ denotes the poisoned sample after alignment; after this, the superscript of poisoned samples is uniformly denoted as $a$. $y_t$ corresponds to $y_t^C$ and $y_t^a$; $\hat{y}_0$ corresponds to $\hat{y}_0^C$ and $\hat{y}_0^a$. $\hat{x}_0$ denotes the final denoised image of the clean sample, while $\hat{x}_0^a$ is ultimately mapped to the target image $T_1$ or $T_2$.

\subsubsection{Regularization Extension: Representation Dispersion Optimization}
Although l-DAE~\cite{D16} performs well in representation learning, we observe that its feature distributions lack sufficient uniformity, which may limit downstream generalization. 
To address this, we introduce a representation dispersion regularization mechanism $L_{\text{disp}}$ ~\cite{D17}, derived from the InfoNCE loss, to encourage uniformity in latent representations.

As shown in Figure~\ref{fig:one}, we extract latent representations 
$y_t^{(l)} \in \mathbb{R}^{B \times D}$ from the $l$-th Transformer block of the DiT model, where $B$ denotes the batch size and $D$ the feature dimension. 
The dispersion loss is defined as:
\begin{equation}
L_{\text{disp}} = \log \mathbb{E}_{i,j} 
\left[
\exp \left(
-\frac{\| y_{t,i}^{(l)} - y_{t,j}^{(l)} \|_2^2}{\tau}
\right)
\right],
\end{equation}
where $\|\cdot\|_2$ represents the Euclidean distance in feature space 
and $\tau$ is a temperature parameter controlling the strength of distribution uniformity. 
This loss is computed over each batch to maximize pairwise feature separation, thus promoting a more uniform latent feature distribution. 
Unlike contrastive learning objectives that rely on explicit sample pairing, this mechanism improves representation uniformity in a self-supervised manner without interfering with the original denoising training objective.

\subsubsection{Unified Optimization Framework}

Building on the above components, we propose a unified Regularized Self-Supervised Diffusion (RSSD) framework. 
This framework integrates PCA-space diffusion denoising with representation dispersion regularization, enhancing representation learning while maintaining generation quality. 
It achieves a balance between \emph{alignment} and \emph{uniformity} without relying on complex data augmentation.
The overall optimization objective minimizes a weighted combination of self-supervised and dispersion losses:
\begin{equation}
L_{\text{RSSD}} = L_{\text{SSL}} + \lambda_{\text{disp}} \cdot L_{\text{disp}},
\end{equation}
where $L_{\text{SSL}}$ is the core self-supervised loss, 
and $\lambda_{\text{disp}}$ controls the contribution of the dispersion term.
The self-supervised loss is defined as:
\begin{equation}
L_{\text{SSL}} = \mathbb{E}_{(x_0, t)} 
\left[
\lambda_t \sum_{j=1}^{N}\sum_{i=1}^{D} w_i \cdot (r_i^{(j)})^2
\right],
\end{equation}
where $r_i^{(j)} = V^\top (x_0^{(j)} - \hat{x}_0^{(j)})$ denotes the residual projection 
of the $j$-th image patch in PCA space, 
$D = P \times P \times C = 768$ is the PCA dimensionality, 
$w_i$ is the weighting coefficient 
(set to 1.0 for the first 48 dimensions and 0.1 for the remaining ones), 
and $\lambda_t = 1 / (1 + \sigma_t^2)$ is the time-dependent noise weighting factor.

\subsection{BadRSSD: Backdoor Attacks on RSSD}
\label{sec3.3}

Building on the RSSD framework, we identify a new attack surface in the model’s representation learning and propose BadRSSD, a stealthy backdoor attack targeting RSSD’s internal representations. 
Unlike conventional backdoors that alter generated outputs, BadRSSD implants a trigger by aligning poisoned samples with a chosen target in PCA latent space, forcing their semantic embeddings to match the target while preserving the model’s original performance. 
By exploiting RSSD’s PCA-latent pipeline, BadRSSD avoids complex multi-step Markov chain manipulations used in DDPM attacks and constructs a concise end-to-end attack chain from latent representation to pixel reconstruction. 
As illustrated in Figure \ref{fig:one}, clean samples follow the standard RSSD denoising–reconstruction path, whereas poisoned samples undergo PCA-space alignment to covertly control model behavior.

\subsubsection{Poisoned Sample Generation and Backdoor Alignment}

\textbf{Poisoned sample generation.} A random noise trigger \(e_i\) is injected into the lower-right \(n\times n\) region of the original image \(x_0\) to create a poisoned sample \(x_p = x_0 + M \odot T\), where \(M\) is a binary mask that equals 1 inside the trigger region and 0 elsewhere.\\
\textbf{PCA-space alignment.} The core mechanism of BadRSSD is semantic alignment in the PCA latent space. After PCA encoding, the poisoned sample \(P_i\) is adjusted so that its encoded representation \(Z_0^{P_i}\) is shifted by \(\Delta z=Z_0^{T_i}-Z_0^{P_i}\) to match the target representation \(Z_0^{T_i}\), i.e., \(Z_0^{a_i} = Z_0^{P_i} + \Delta z = Z_0^{T_i}\). This operation directly modifies the poisoned sample's semantic identity in RSSD's PCA space, enforcing alignment with the target while preserving model utility and stealth.

\subsubsection{Core Loss Function Design}

BadRSSD adopts a conditional training strategy that dynamically adjusts the optimization objectives based on sample type.
For \textbf{clean samples}, the model follows the standard self-supervised representation learning objective:
\begin{equation}
L_{\text{clean}} = L_{\text{RSSD}} = L_{\text{SSL}} + \lambda_{\text{disp}} \cdot L_{\text{disp}}.
\end{equation}
For \textbf{poisoned samples}, the model is optimized under a backdoor loss that enforces semantic alignment between the trigger and the target image. 
The loss function is defined as:
\begin{equation}
\begin{split}
\theta^* 
&= \arg \min_{\theta} L_{\text{backdoor}} \\
&= \alpha_1 L_{\text{PCA\_TR}}(\theta) 
+ \alpha_2 L_{\text{img\_rec}}(\theta) 
+ \alpha_3 L_{\text{disp}}(\theta),
\end{split}
\end{equation}\\
where $\alpha_1$, $\alpha_2$, and $\alpha_3$ are hyperparameters that balance the contributions of each term. 
These three core loss components jointly optimize representation consistency, reconstruction fidelity, and dispersion regularization.\\
\textbf{PCA trajectory dual alignment loss.}
The PCA trajectory dual alignment loss ensures that poisoned samples remain semantically aligned with their target images throughout the diffusion process. 
It consists of two components: 
(1) \textit{static alignment $L_\text{PCA}$}, which enforces similarity between the initial PCA-encoded features $Z_0^{a_i}$ and $Z_0^{T_i}$; 
and (2) \textit{dynamic consistency $L_\text{TR}$}, which maintains alignment of their trajectories over time for $Z_t^{a_i}$ and $Z_t^{T_i}$. 
The loss is defined as:
\begin{equation}
\begin{split}
L_{\text{PCA\_TR}} 
&= \mathbb{E}_{(x_p, x_{T_i}),t} \Big[
\lambda_t \underbrace{\sum_{j=1}^{D} w_j (Z_{0,j}^{a_i} - Z_{0,j}^{T_i})^2}_{L_\text{PCA}} \\
&\quad + \underbrace{\beta \| Z_t^{a_i} - Z_t^{T_i} \|_2^2}_{L_\text{TR}}
\Big],
\end{split}
\end{equation}\\
where $E_{(x_p,x_{T_i}),t}$ denotes the expectation over poisoned samples $x_p$, target samples $x_{T_i}$, and diffusion step $t$;
$\lambda_t = 1 / (1 + \sigma_t^2)$ is the noise-adaptive weighting factor;
$D = 768$ is the PCA dimensionality;
$w_j$ represents the PCA dimension weight (1.0 for the first 48 dimensions and 0.1 for the rest);
and $\beta$ balances static and dynamic alignment to ensure consistent semantic trajectories in PCA space.\\
\textbf{Image Reconstruction Loss.}
The image reconstruction loss ensures that, after denoising by the DiT model and decoding via the VAE, poisoned samples can accurately reconstruct the target image at the pixel level. 
As shown in Figure~\ref{fig:one}, this loss directly constrains the final image output, 
compensating for the limited fine-grained control of PCA-space alignment. 
The loss is defined as:
\begin{equation}
L_{\text{img\_rec}} = \mathbb{E}_{(x_p, x_T),t}
\left[
\left\|
x_0^{T_i} - 
\underbrace{\text{VAE}_{\text{decode}}\left(\text{DiT}(y_t^{a}, t)\right)}_{ViT_t}
\right\|_2^2
\right],
\end{equation}
where $x_0^{T_i}$ denotes the target image representation and $ViT_t$ represents the DiT model’s predicted denoising output at timestep $t$, which is decoded back to image space by the VAE. 
The $\|\cdot\|_2$ term measures pixel-level reconstruction accuracy.\\
\textbf{Representation Dispersion Loss.}
Following the RSSD framework, this loss enhances attack stealth by promoting uniform feature distribution in latent space. 
Together with PCA-space alignment, it balances compactness and dispersion, improving attack success while maintaining the model’s normal representation learning ability and robustness against anomaly detection.
\section{Performance Evaluation}
\label{sec:Experimental}

\begin{table*}[ht]
\small
\centering
\resizebox{\textwidth}{!}{%
\begin{tabular}{@{}c|c|ccccccccccccc@{}}
\hline
\textbf{ Pre-training} & \textbf{ Downstream} & \textbf{ Benign} & \multicolumn{2}{c}{\textbf{ BadEncoder}} & \multicolumn{2}{c}{\textbf{SSLBKD}} & \multicolumn{2}{c}{\textbf{Ours}} & \multicolumn{2}{c}{\textbf{BadDiffusion}} & \multicolumn{2}{c}{\textbf{TrojDiff}} & \multicolumn{2}{c}{\textbf{Ours}} \\
\cline{3-15}
\textbf{Dataset} & \textbf{Dataset} & CA$\uparrow$ & BA$\uparrow$ & ASR$\uparrow$ & BA$\uparrow$ & ASR$\uparrow$ & BA$\uparrow$ & ASR$\uparrow$ & FID$\downarrow$ & MSE$\downarrow$ &FID$\downarrow$ & MSE$\downarrow$ & FID$\downarrow$ & MSE$\downarrow$  \\

\hline
CelebA-HQ   & CelebA-HQ & \textbf{85.23} & 83.52 & 86.78 & 54.26 & 62.57 & 83.96 & \textbf{92.63} & 46.15 & 0.2124 & 47.06 & 0.2376 & \textbf{38.26} & \textbf{0.1625} \\
\hline
CIFAR-100   & CIFAR-10 & 82.36 & 81.45 & 73.86 & 50.65 & 53.72 & \textbf{83.65} & \textbf{91.26} & 42.86 & 0.1628 & 43.25 & 0.1726 & \textbf{36.12} & \textbf{0.0821} \\
\hline
              & CelebA-HQ & \textbf{85.57} & 74.56 & 76.42 & 48.75 & 52.53 & 84.92 & \textbf{94.67} & 47.26 & 0.1732 & 50.12 & 0.1935 & \textbf{38.52} & \textbf{0.1209} \\
ImageNet & CIFAR-10 & \textbf{73.56} & 70.26 & 78.92 & 46.63 & 55.72 & 72.14 & \textbf{86.75} & 53.82 & 0.2028 & 55.19 & 0.2268 & \textbf{40.18} & \textbf{0.1526} \\
             & CIFAR-100 & \textbf{72.24} & 68.55 & 80.16 & 49.46 & 57.23 & 71.94 & \textbf{87.12} & 55.27 & 0.2236 & 57.38 & 0.2435 & \textbf{43.26} & \textbf{0.1815} \\
\hline
\end{tabular}
}
 \caption{Camparison of attack performance on different datasets. The best result are  \textbf{highlighted}.}
 \label{tab:one}
\end{table*}

\begin{table}[htbp]
\small
\centering
\setlength{\tabcolsep}{2.5pt}
\begin{tabular}{lcccccc}
\toprule
\textbf{Samplers} & \textbf{ASR} & \textbf{FID} & \textbf{MSE} & \textbf{Steps} & \textbf{Time (s)} & \textbf{ASR-std} \\
\midrule
DDPM & 92.37 & 40.12 & 0.1263 & 1000 & 15.2 & 2.1 \\
DDIM & 72.18 & 52.38 & 0.2514 & 50 & 4.2 & 0.5 \\
DPM-solver & 87.53 & 41.84 & 0.1347 & 20 & 3.8 & 1.2 \\
Euler & 78.96 & 49.16 & 0.2281 & 100 & 2.1 & 1.8 \\
\bottomrule
\end{tabular}
\caption{Comparison of different samplers.}
\label{tab:two}
\end{table}

To demonstrate the effectiveness and robustness of our method, we implemented BadRSSD using PyTorch and compared its performance with existing state-of-the-art and highly relevant backdoor attack approaches. All experiments were conducted on an NVIDIA A100 GPU with 80GB of memory, and each experiment was repeated five times under independent runs, with their average results reported. We designed comprehensive experiments to address the following three research questions:\\
\textbf{RQ1 (Effectiveness of BadRSSD)}: Can BadRSSD succes-sfully inject a backdoor into RSSD?\\
\textbf{RQ2 (Visual Analysis of BadRSSD’s Stability)}: Can BadRSSD maintain stable performance across different settings, as analyzed through visualization?\\
\textbf{RQ3 (Robustness of BadRSSD)}: Can BadRSSD effectively resist existing defense methods?
\subsection{Experimental Setup}
\textbf{Datasets.} We conducted experiments on four datasets: CIFAR-10~\cite{CIFAR10}, CelebA-HQ~\cite{D48}, CIFAR-100~\cite{CIFAR10}, and ImageNet~\cite{Imagenet}. These datasets were selected for their diversity in scale, semantic domain, and resolution (Table~\ref{tab:one}) to thoroughly assess the attack's effectiveness and generalization across varied conditions. Unless stated otherwise, ImageNet served as the pre-training dataset and CelebA-HQ as the downstream dataset. Additional dataset details are provided in Supplementary Sec. 1.\\
\textbf{Evaluation Metrics.} Following established SSL backdoor evaluation protocols~\cite{Badencoder,B46}, we employ three core metrics: Clean Accuracy (CA), Attack Success Rate (ASR), and Backdoor Accuracy (BA). An effective attack must maximize ASR while maintaining high BA. For diffusion model attacks~\cite{H22,Trojdiff}, performance is quantified using Fréchet Inception Distance (FID)~\cite{FID}, Mean Squared Error (MSE), and Structural Similarity (SSIM)~\cite{SSIM}. A successful attack achieves low FID (preserving generation quality), minimized MSE, and maximized SSIM, ensuring triggered samples closely match target images. Detailed methodology is provided in Supplementary Sec. 2.\\
\textbf{Baseline.} For comprehensive evaluation, BadRSSD is compared against four state-of-the-art image-patch backdoor attacks targeting two model categories: \textbf{SSL Backdoor Attacks} (BadEncoder~\cite{Badencoder}, SSLBKD~\cite{Badencoder}) and \textbf{Diffusion Model Backdoor Attacks} (BadDiffusion~\cite{H22}, TrojDiff~\cite{Trojdiff}). All methods are implemented under identical DiT-L/2 architectures with consistent triggers and poisoning strategies. SSL attacks are measured by BA/ASR, while diffusion attacks are evaluated via FID/MSE.\\
\begin{figure}[ht]
  \centering
  \includegraphics[width=0.75\linewidth]{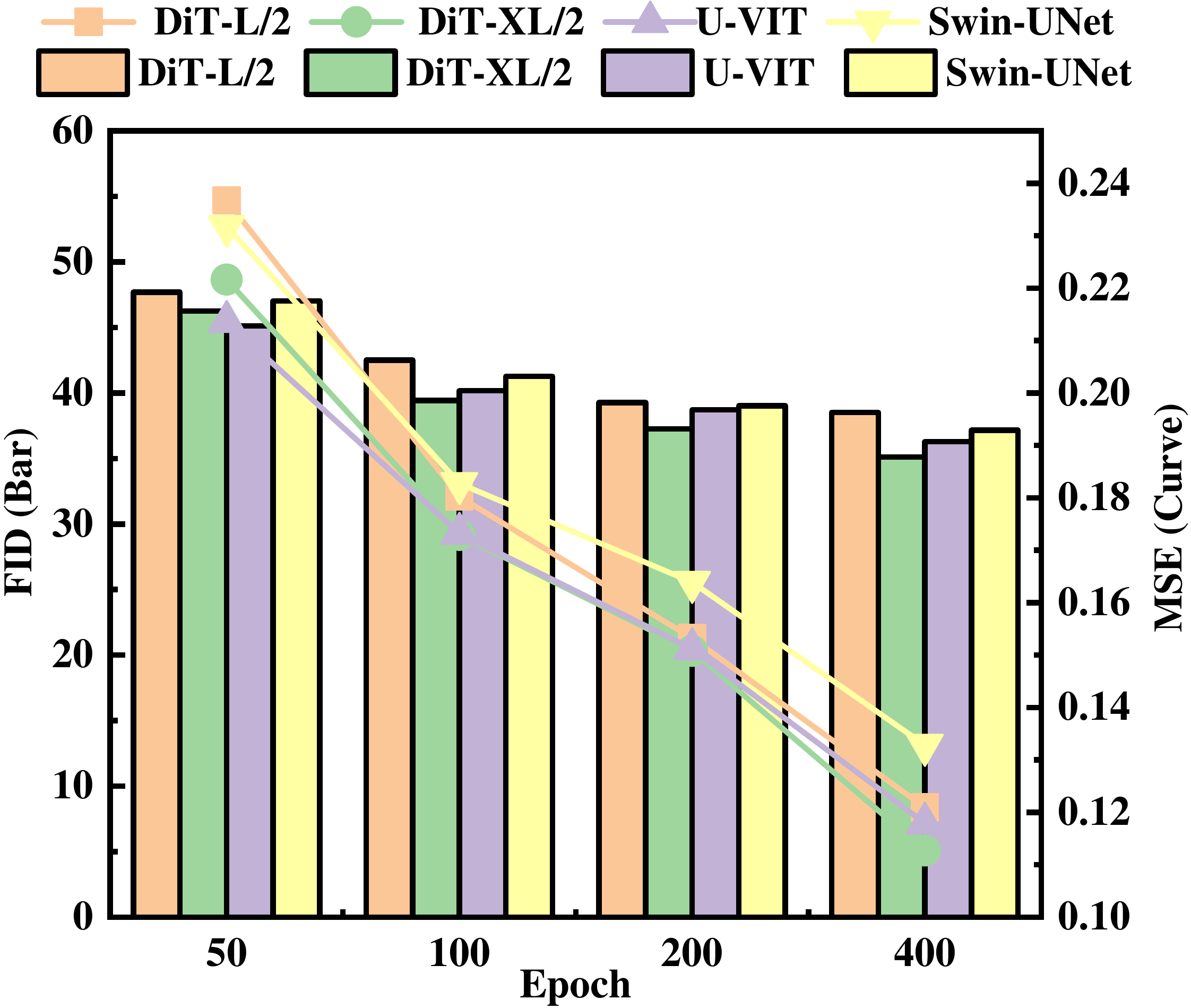}
   \caption{Experimental results for different Transformer architectures. \textbf{Clean-FID (bars) and Backdoor-MSE (curves)}.}
   \label{fig:two}
\end{figure}
\textbf{Implementation Details.}
We employ the standard DiT-L/2-SSL model, a self-supervised variant of DiT-L~\cite{S19} using a half-depth ViT-L architecture~\cite{ViT} with 12 encoder/decoder layers (total 24 blocks) and MLP ratio 1/4. Following~\cite{D16,D17,S19}, RSSD base training uses Adam~\cite{Adam} (lr=1e-4, batch 2048, 400 epochs). BadRSSD attacks apply Adam (lr=1e-4, batch 32, 100/400 epochs) with DPM-solver sampling~\cite{Dpm-solver} (20 steps), linear noise scheduling ($\alpha=0.1$), and $5\%$ poisoning rate. Additional implementation details in Supplementary Sec. 3.

\subsection{Effectiveness Evaluation (RQ1)}
\textbf{Effectiveness comparison with SOTA attack methods.} BadRSSD's effectiveness is evaluated through two dimensions: SSL backdoor performance and diffusion attack quality. As shown in Table~\ref{tab:one}, BadRSSD consistently outperforms baselines across all scenarios. For SSL backdoor evaluation, BadRSSD achieves the highest ASR (exceeds $86\%$) while maintaining benign accuracy, significantly surpassing BadEncoder~\cite{Badencoder} and SSLBKD~\cite{B46}. In diffusion quality assessment, BadRSSD attains optimal FID (36.12) and MSE (0.0821), demonstrating precise backdoor control with superior generation fidelity compared to BadDiffusion~\cite{H22} and TrojDiff~\cite{Trojdiff}. These results collectively validate BadRSSD's robustness and generalization capability.\\
\textbf{Effectiveness on different Transformer architectures.}
BadRSSD's architectural generalization is validated across four Transformer backbones (DiT-L/2~\cite{S19}, DiT-XL/2~\cite{S19}, U-ViT~\cite{U-ViT}, and Swin-UNet~\cite{Swin-unet}; see Supplementary Sec. 4 for architectural details). As Figure~\ref{fig:two} shows, all models achieve decreasing FID/MSE with extended training (50→400 iterations), confirming improved generation quality and attack precision. DiT-XL/2 performs best (FID 35.12, MSE 0.1127 at 400 iterations), followed by U-ViT (FID 36.28, MSE 0.1179). The consistent performance across architectures demonstrates BadRSSD's universal applicability to transformer-based diffusion models.\\
\textbf{Effectiveness on different sampler.}
BadRSSD's sampler comparison (DDPM~\cite{DDPM}, DDIM~\cite{DDIM}, DPM-solver~\cite{Dpm-solver}, and Euler~\cite{E53}) (Table~\ref{tab:two}) reveals distinct performance trade-offs. DDPM achieves peak ASR ($92.37\%$) and optimal FID/MSE (40.12/0.1263) at 1000 steps but requires 15.2s sampling. DPM-solver attains competitive performance ($87.53\%$ ASR, 41.84 FID, 0.1347 MSE) in merely 20 steps (3.8s), offering the best efficiency-effectiveness balance. DDIM shows highest stability (ASR-std 0.5) at 50 steps but with compromised attack quality. Consequently, DPM-solver is selected as BadRSSD's primary sampler. Here, ASR-std denotes the standard deviation of the Attack Success Rate, where a smaller value indicates greater stability across different random seeds.

\begin{table*}[htbp]
\small
\centering
\includegraphics[width=\textwidth]{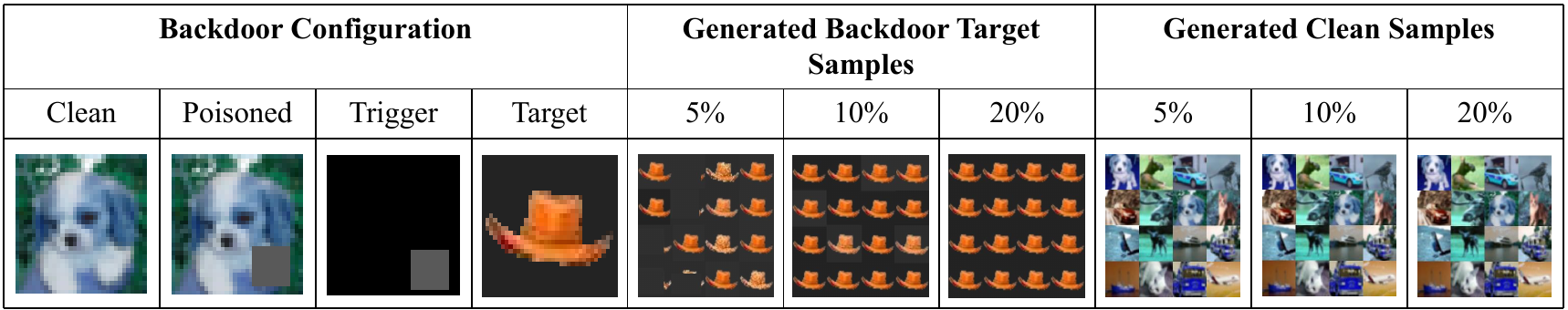}
\vspace{-1.5em} 
\caption{Visualized examples of BadRSSD on low-resolution CIFAR-10 with trigger Grey Box \& target hat and without triggers at different poison rates.~(Triggers are Gray Box. target categories: $T_1$ (orange hat): CIFAR-10/100; $T_2$ (cartoon girl): CelebA-HQ/ImageNet.)}
\label{tab:three}
\end{table*}

\begin{figure*}[htbp]
\small
\centering
\begin{subfigure}{0.30\textwidth}
    \centering
    \includegraphics[width=\textwidth, height=4cm, keepaspectratio]{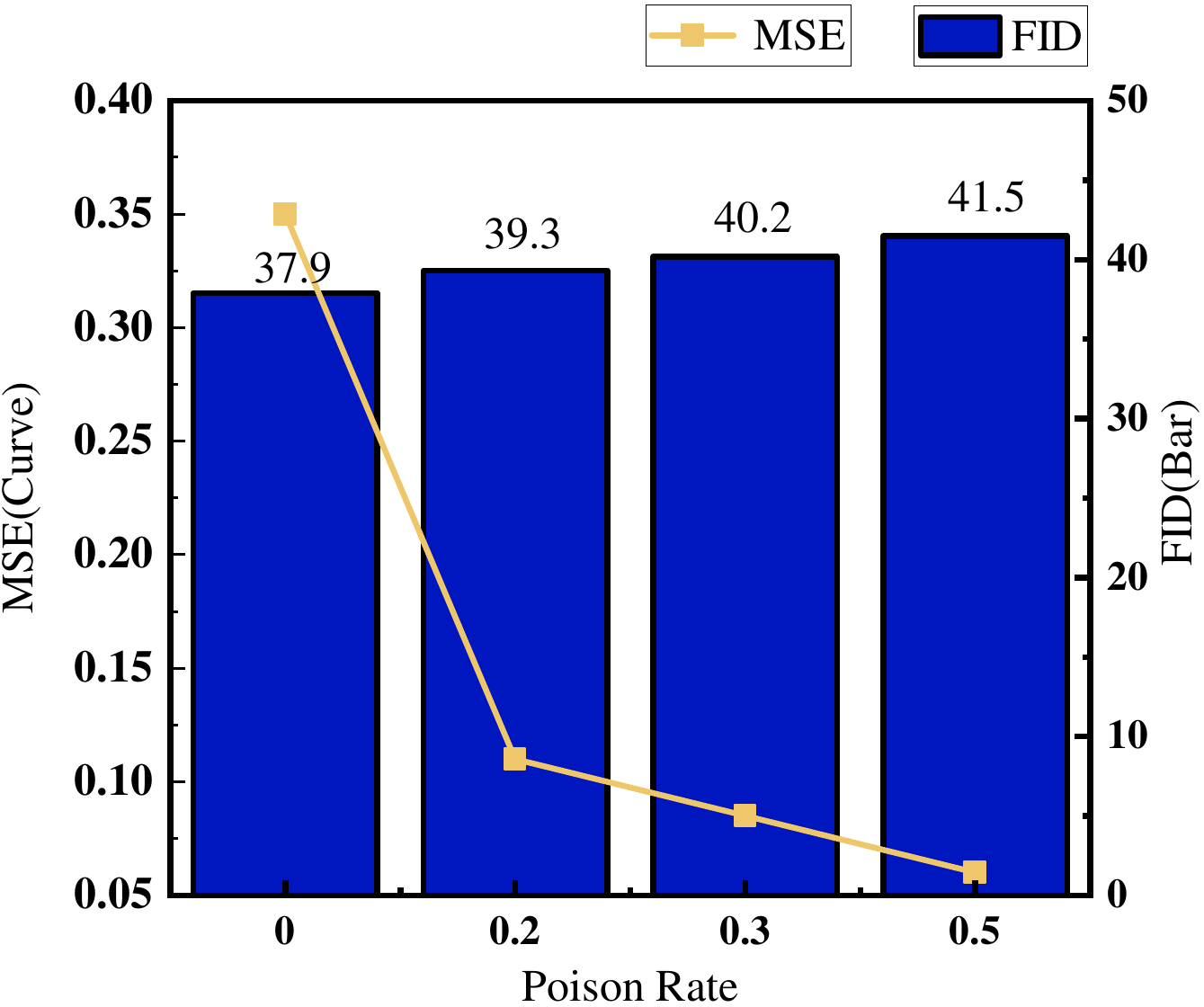}
    \caption{FID (bars) and MSE (curves)}
    \label{fig:three1}
\end{subfigure}
\hspace{0.2cm} 
\begin{subfigure}{0.60\textwidth}
    \centering
    \includegraphics[width=\textwidth, height=4.5cm, keepaspectratio]{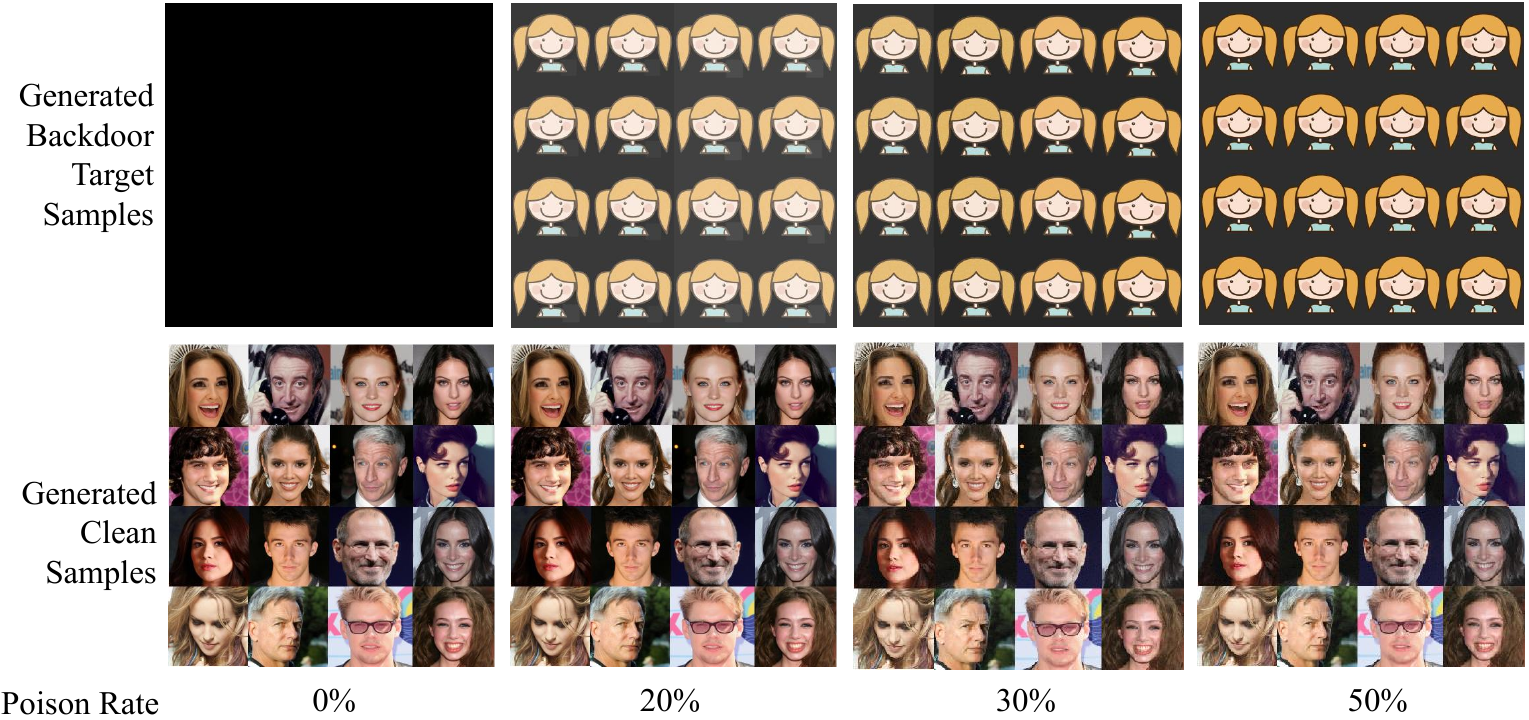}
    \caption{Visual examples}
    \label{fig:three2}
\end{subfigure}
\caption{ On high-resolution CelebA-HQ, BadRSSD utilizes a gray-scale box trigger with a target cartoon girl across varying poisoning rates. Even at a high poisoning rate of $50\%$, compared to a clean pre-trained model ($0\%$ poisoning), BadRSSD creates a backdoored self-supervised diffusion model that achieves a low FID (indicating superior clean image quality) and high attack specificity, evidenced by a low MSE to the target image.~Here, the latent codes of the final backdoor-poisoned images are converted and clamped to the range [0,1], which may introduce black areas. The results in Table 3 and Figure 4 are likewise processed.}
\label{fig:three}
\end{figure*}


\begin{figure}[h]
  \centering
  \includegraphics[width=0.8\linewidth]{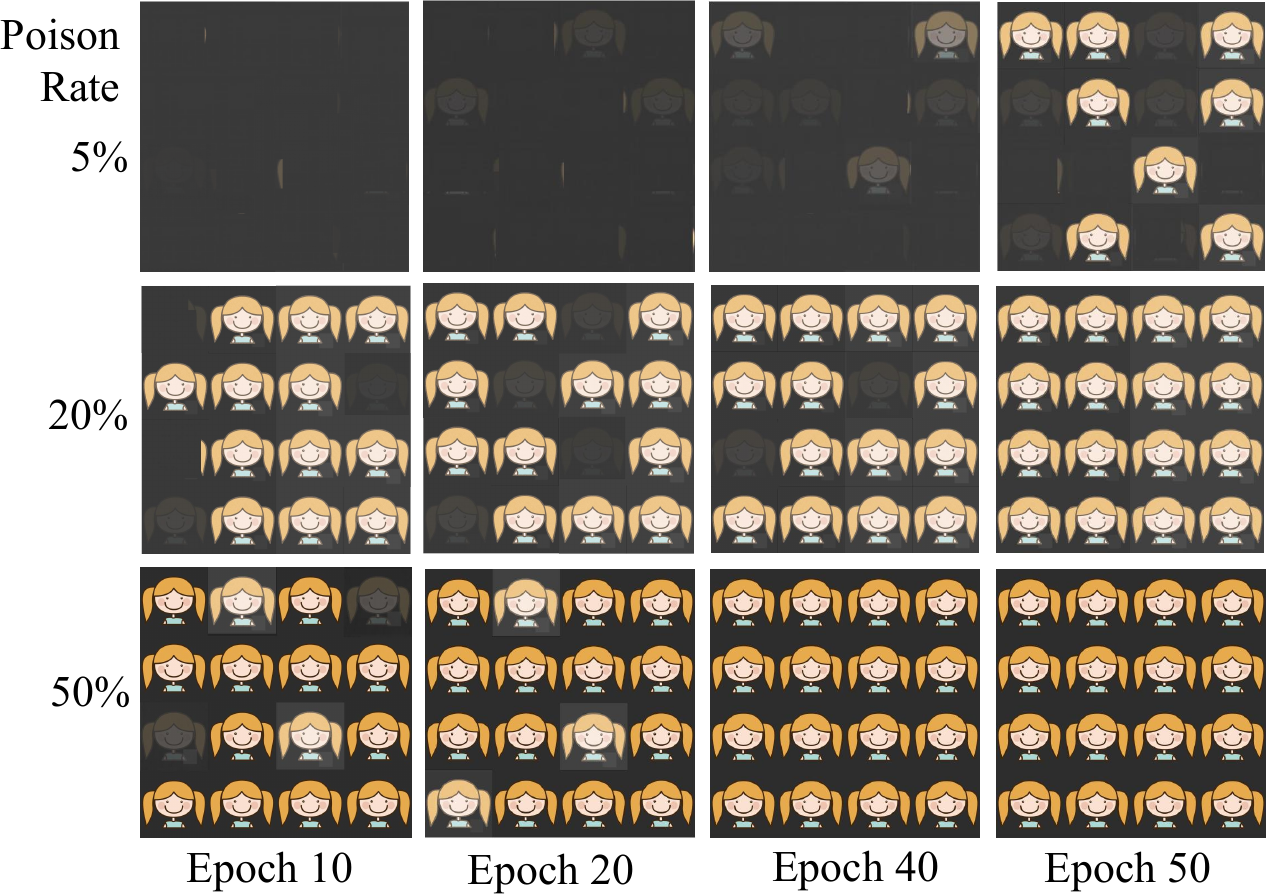}
   \caption{Visualized samples of backdoor objectives synthesized by BadRSSD  on ImageNet with trigger Grey Box \& target cartoon girl across different training epochs. }
   \label{fig:four}
\end{figure}
\begin{table*}[ht]
\small
\centering
\resizebox{\textwidth}{!}{%
\begin{tabular}{lccccccccc}
\toprule
Methods & PDD(Clean) & PDD(Poison) & AUROC & TPR@1\%FPR & Detection Pass Rate(\%) & ASR-before(\%) & ASR-after(\%) & AFID \\
\midrule
BadDiffusion & 0.22 & 0.67 & 0.95 & 87.12 & 26.23 & 71.02 & 8.26 & -0.58 \\
TrojDiff & 0.28 & 0.72 & 0.92 & 85.46 & 32.17 & 73.18 & 9.45 & 0.12 \\
Ours & 0.31 & 0.43 & 0.58 & 8.72 & 85.68 & 94.67 & 92.57 & 4.63 \\
\bottomrule
\end{tabular}
}
\caption{Experimental results of DisDet.}
\label{tab:four}
\end{table*}
\begin{table*}[htbp]
\small
\centering
\resizebox{\textwidth}{!}{%
\begin{tabular}{lcccccccccc}
\toprule
\multirow{2}{*}{Poison Rate} & \multicolumn{5}{c}{BadDiffusion} & \multicolumn{5}{c}{Ours} \\
\cmidrule(lr){2-6} \cmidrule(lr){7-11}
& Detected(\%) & ASR-before(\%) & ASR-after(\%) & ASSIM & AFID & Detected(\%) & ASR-before(\%) & ASR-after(\%) & ASSIM & AFID \\
\midrule
0.05 & 82.06 & 71.02 & 4.02 & -0.8 & -0.23 & 5.12 & 94.67 & 93.72 & 0 & -0.89 \\
0.1 & 89.72 & 76.52 & 4.16 & -0.9 & 0.41 & 8.26 & 96.64 & 95.05 & 0 & 1.2 \\
0.2 & 94.53 & 82.32 & 5.13 & -1 & -0.15 & 12.15 & 97.92 & 96.69 & 0.01 & 2.7 \\
0.3 & 99.99 & 84.51 & 5.16 & -1 & 0.32 & 16.36 & 99.16 & 97.47 & -0.01 & 4.6 \\
0.4 & 100.00 & 89.96 & 5.25 & -1 & 0.43 & 17.12 & 99.99 & 98.02 & -0.01 & 6.5 \\
0.5 & 100.00 & 92.65 & 6.07 & -1 & 0.52 & 18.08 & 99.99 & 98.65 & -0.01 & 8.5 \\
\bottomrule
\end{tabular}
}
\caption{Experimental results of Elijah.}
\label{tab:five}
\end{table*}

\subsection{Visual Stability Analysis (RQ2)}
A multi-dimensional stability evaluation across diverse resolutions, training configurations, and poisoning rates was conducted to verify the effectiveness and stealth of the BadRSSD attack.\\
\textbf{Attack Effectiveness and Model Integrity on Low-Resolution Datasets.} On the low-resolution CIFAR-10 (32×32, Table~\ref{tab:three}), backdoor success escalates with poisoning rates: partial at $5\%$, complete at $10–20\%$, confirming high specificity. Meanwhile, clean sample quality remains unaffected across all rates, balancing high utility and specificity.\\
\textbf{Quantitative Robustness and Attack Specificity on High-Resolution Datasets.}
On high-resolution CelebA-HQ (256×256) (Figure~\ref{fig:three}), FID shows minimal increa-se (37.9→41.5, +$9.5\%$) with rising poisoning rates ($0\%$→$50\%$), confirming high utility. Concurrently, MSE drops to 0.03, achieving high specificity. Visual results confirm stable clean generation and perfect target reproduction above $20\%$ poisoning, demonstrating robust performance.
\textbf{Training Dynamics and Backdoor Persistence Analysis.} Training dynamics on ImageNet (256×256, Figure~\ref{fig:four}) confirm the backdoor’s temporal stability. At 5\% poisoning, target synthesis emerges only partially by Epoch 50; at 20\%, it stabilizes by Epoch 40; and at 50\%, full stability is reached as early as Epoch 20. Once established, the backdoor remains persistent, ensuring sustained high specificity throughout training.


\begin{table}[tb]
\small
    \centering
    \begin{tabular}{c| c c c c}
        \toprule
        \multirow{2}{*}{Performance} & \multicolumn{2}{c}{BadDiffusion} & \multicolumn{2}{c}{Ours}\\
         \cmidrule(lr){2-5}
        Metrics & $\eta$ = 1 & $\eta$ = 0.1 & $\eta$ = 1 & $\eta$ = 0.1\\
        \midrule
        ASR\_before & 71.02 & 68.19 & 94.67 & 92.51\\
        \hline
        ASR\_after & 8.12 & 7.65 & 92.35 & 91.76\\
        \hline
        $\Delta$FID & -0.36 & 0.52 & 8.16 & 4.52\\
        \hline                         
        TPR(\%) & 85.12 & 76.56 & 6.74 & 4.28\\    
        \hline                        
        TNR(\%) & 95.67 & 96.12 & 97.26 & 96.89\\ 
        \hline                          
        $\|\mathbf{r} - \mathbf{r}_0\|_2$ & 0.18 & 0.26 & 0.82 & 0.75\\ 
        \bottomrule
    \end{tabular}
    \caption{Experimental results of TERD.}
    \label{tab:six}
\end{table}

\subsection{Robustness Evaluation (RQ3)}
\textbf{Resistance to DisDet.}
DisDet~\cite{Disdet} quantifies clean/poiso-ned sample disparity via distribution discrepancy (PDD). As shown in Table~\ref{tab:four}, BadDiffusion/TrojDiff exhibit significant PDD separation, yielding high detection efficacy (AUROC$\approx$0.92--0.95, TPR@1\%FPR$\approx$85--87\%) and ASR reduction to 8-9\% post-mitigation. In contrast, BadRSSD shows minimal PDD divergence (0.31 vs. 0.43), resulting in random-level AUROC (0.58), low TPR@1\%FPR (8.72\%), and high evasion rate (85.68\%). Its ASR remains stable (94.67$\to$92.57) with minor $\Delta$FID increase (+4.63), confirming detection failure. This stems from BadRSSD's distribution stability enforced by $L_{\text{disp}}$, where backdoor effects manifest primarily in PCA subspace and denoising stages, becoming statistically indistinguishable at low poisoning rates.\\
\textbf{Resistance to Elijah.}
Elijah~\cite{Elijah} combats backdoors by inverting triggers and pruning correlated neurons. It effecti-vely defends against BadDiffusion, achieving $\approx$100\% detection and reducing ASR to 4--6\%(Table~\ref{tab:five}). In contr-ast, against BadRSSD, Elijah attains only $\approx$5--18\% detection, leaving ASR nearly unchanged ($\Delta\text{ASR} \leq 3\%$) and $\Delta\text{SSIM} \approx 0$, with only $\Delta\text{FID}$ increasing with poisoning rate. This failure arises because BadRSSD's trigger is a subtle, non-local perturbation aligned with target semantics in PCA space, resisting pixel-space inversion. Moreover, its backdoor pathway is temporally and spatially dispersed, preventing effective neuron pruning.\\
\textbf{Resistance to TERD.}
TERD~\cite{TERD} detects backdoors by inverting structured triggers in pixel space. As Table~\ref{tab:six} shows, it effectively mitigates BadDiffusion (ASR: 71\%$\to$8\%) with high TPR (85\%) and accurate trigger inversion ($\|\mathbf{r} - \mathbf{r}_0\|_2$$\approx 0.18$). However, against BadRSSD, TERD achieves near-zero TPR (7\%) while maintaining high TNR (97\%), with trigger inversion failing ($\|\mathbf{r} - \mathbf{r}_0\|_2$$\approx 0.82$) and ASR remaining unchanged (95\%$\to$ 92\%). This failure stems from BadRSSD's PCA-driven semantic alignment and dispersed temporal activation, which create a fundamental mismatch with pixel-space inversion methods.
\textbf{Experimental details and metrics are in Supplementary Sec. 5.}

\subsection{Ablation Study}
Ablation studies validate the necessity of each loss component in BadRSSD (Table~\ref{tab:seven}). Removing $L_\text{PCA\_TR}$ reduces ASR from 94.67\% to 62.18\% and increases FID from 38.52 to 61.23, confirming its critical role in attack execution. Eliminating $L_\text{img\_rec}$ degrades reconstruction fidelity (MSE: 0.1209$\to$0.2689, SSIM: 0.823$\to$0.658), while removing $L_\text{disp}$ slightly raises ASR to 95.16\% but increases FID to 49.52, underscoring its essential role in maintaining generation quality and stealth. These results collectively demonstrate the complementary functions of the three losses. (\enquote{All losses} uses all three losses, while \enquote{No $L_{i}$} excludes the $i$-th loss.)
\begin{table}[ht]
\small
\centering
\begin{tabular}{lccccc}
\toprule
& ASR$\uparrow$ & FID$\downarrow$ & MSE$\downarrow$ & SSIM$\uparrow$ \\
\midrule
All losses & 94.67 & 38.52 & 0.1209 & 0.823 \\
No $L_\text{PCA\_TR}$  & 62.18 & 61.23 & 0.3825 & 0.521 \\
No $L_\text{img\_rec}$ & 82.15 & 48.67 & 0.2689 & 0.658 \\
No $L_\text{disp}$ & 95.16 & 49.52 & 0.1789 & 0.768 \\
\bottomrule
\end{tabular}
\caption{Performance of Ablation Studies.}
\label{tab:seven}
\end{table}
\section{Conclusion}
\label{sec:Conclusion}
This paper presents Regularized Self-Supervised Diffusion (RSSD) and its backdoor variant BadRSSD. RSSD unifies generative and representational learning, while BadRSSD employs PCA-space backdoor alignment with conditional triple-loss to achieve potent, stealthy attacks—revealing serious security implications of representation-layer backdoors. Future work will focus on developing corresponding defenses and security standards.

{
    \small
    \bibliographystyle{ieeenat_fullname}
    \bibliography{main}

@String(ECCV= {Eur. Conf. Comput. Vis.})

@String(ICLR = {Int. Conf. Learn. Represent.})

@String(AAAI = {AAAI})

@String(ECCV  = {ECCV})

@String(ICLR  = {ICLR})

@article{DDPM,
  title={Denoising diffusion probabilistic models},
  author={Ho, Jonathan and Jain, Ajay and Abbeel, Pieter},
  journal={Advances in neural information processing systems},
  volume={33},
  pages={6840--6851},
  year={2020}
}

@article{DMBG,
  title={Diffusion models beat gans on image synthesis},
  author={Dhariwal, Prafulla and Nichol, Alexander},
  journal={Advances in neural information processing systems},
  volume={34},
  pages={8780--8794},
  year={2021}
}

@inproceedings{Diffusion-4K,
  title={Diffusion-4k: Ultra-high-resolution image synthesis with latent diffusion models},
  author={Zhang, Jinjin and Huang, Qiuyu and Liu, Junjie and Guo, Xiefan and Huang, Di},
  booktitle={Proceedings of the Computer Vision and Pattern Recognition Conference},
  pages={23464--23473},
  year={2025}
}

@inproceedings{EmoReg,
  title={EmoReg: Directional Latent Vector Modeling for Emotional Intensity Regularization in Diffusion-based Voice Conversion},
  author={Gudmalwar, Ashishkumar Prabhakar and Biyani, Ishan Darshan and Shah, Nirmesh J and Wasnik, Pankaj and Shah, Rajiv Ratn},
  booktitle={Proceedings of the AAAI Conference on Artificial Intelligence},
  volume={39},
  number={22},
  pages={23960--23968},
  year={2025}
}

@inproceedings{DIFFVSGG,
  title={DIFFVSGG: Diffusion-Driven Online Video Scene Graph Generation},
  author={Chen, Mu and Li, Liulei and Wang, Wenguan and Yang, Yi},
  booktitle={Proceedings of the Computer Vision and Pattern Recognition Conference},
  pages={29161--29172},
  year={2025}
}

@article{DiffChar,
  title={DiffChar: A Fast Conditional Diffusion Model For Air-Writing Chinese Character Generation},
  author={Zhao, Weixi and Wu, Meiqi and Wang, Weiqing},
  journal={Pattern Recognition},
  pages={112307},
  year={2025},
  publisher={Elsevier}
}

@inproceedings{D44,
  title={Denoising diffusion autoencoders are unified self-supervised learners},
  author={Xiang, Weilai and Yang, Hongyu and Huang, Di and Wang, Yunhong},
  booktitle={Proceedings of the IEEE/CVF International Conference on Computer Vision},
  pages={15802--15812},
  year={2023}
}

@article{D16,
  title={Deconstructing denoising diffusion models for self-supervised learning},
  author={Chen, Xinlei and Liu, Zhuang and Xie, Saining and He, Kaiming},
  journal={arXiv preprint arXiv:2401.14404},
  year={2024}
}

@article{D17,
  title={Diffuse and Disperse: Image Generation with Representation Regularization},
  author={Wang, Runqian and He, Kaiming},
  journal={arXiv preprint arXiv:2506.09027},
  year={2025}
}

@inproceedings{H22,
  title={How to backdoor diffusion models?},
  author={Chou, Sheng-Yen and Chen, Pin-Yu and Ho, Tsung-Yi},
  booktitle={Proceedings of the IEEE/CVF Conference on Computer Vision and Pattern Recognition},
  pages={4015--4024},
  year={2023}
}

@inproceedings{Trojdiff,
  title={Trojdiff: Trojan attacks on diffusion models with diverse targets},
  author={Chen, Weixin and Song, Dawn and Li, Bo},
  booktitle={Proceedings of the IEEE/CVF Conference on Computer Vision and Pattern Recognition},
  pages={4035--4044},
  year={2023}
}

@article{DDIM,
  title={Denoising diffusion implicit models},
  author={Song, Jiaming and Meng, Chenlin and Ermon, Stefano},
  journal={arXiv preprint arXiv:2010.02502},
  year={2020}
}

@inproceedings{E18,
  title={Extracting and composing robust features with denoising autoencoders},
  author={Vincent, Pascal and Larochelle, Hugo and Bengio, Yoshua and Manzagol, Pierre-Antoine},
  booktitle={Proceedings of the 25th international conference on Machine learning},
  pages={1096--1103},
  year={2008}
}

@inproceedings{S19,
  title={Scalable diffusion models with transformers},
  author={Peebles, William and Xie, Saining},
  booktitle={Proceedings of the IEEE/CVF international conference on computer vision},
  pages={4195--4205},
  year={2023}
}

@article{Villandiffusion,
  title={Villandiffusion: A unified backdoor attack framework for diffusion models},
  author={Chou, Sheng-Yen and Chen, Pin-Yu and Ho, Tsung-Yi},
  journal={Advances in Neural Information Processing Systems},
  volume={36},
  pages={33912--33964},
  year={2023}
}

@inproceedings{Paas,
  title={Personalization as a shortcut for few-shot backdoor attack against text-to-image diffusion models},
  author={Huang, Yihao and Juefei-Xu, Felix and Guo, Qing and Zhang, Jie and Wu, Yutong and Hu, Ming and Li, Tianlin and Pu, Geguang and Liu, Yang},
  booktitle={Proceedings of the AAAI Conference on Artificial Intelligence},
  volume={38},
  number={19},
  pages={21169--21178},
  year={2024}
}

@inproceedings{Rickrolling,
  title={Rickrolling the artist: Injecting backdoors into text encoders for text-to-image synthesis},
  author={Struppek, Lukas and Hintersdorf, Dominik and Kersting, Kristian},
  booktitle={Proceedings of the IEEE/CVF international conference on computer vision},
  pages={4584--4596},
  year={2023}
}

@article{REDEditing,
  title={REDEditing: Relationship-Driven Precise Backdoor Poisoning on Text-to-Image Diffusion Models},
  author={Guo, Chongye and Fu, Jinhu and Fang, Junfeng and Wang, Kun and Feng, Guorui},
  journal={arXiv preprint arXiv:2504.14554},
  year={2025}
}

@inproceedings{TwT,
  title={Silent branding attack: Trigger-free data poisoning attack on text-to-image diffusion models},
  author={Jang, Sangwon and Choi, June Suk and Jo, Jaehyeong and Lee, Kimin and Hwang, Sung Ju},
  booktitle={Proceedings of the Computer Vision and Pattern Recognition Conference},
  pages={8203--8212},
  year={2025}
}

@article{TERD,
  title={TERD: A unified framework for safeguarding diffusion models against backdoors},
  author={Mo, Yichuan and Huang, Hui and Li, Mingjie and Li, Ang and Wang, Yisen},
  journal={arXiv preprint arXiv:2409.05294},
  year={2024}
}

@article{Disdet,
  title={Disdet: Exploring detectability of backdoor attack on diffusion models},
  author={Sui, Yang and Phan, Huy and Xiao, Jinqi and Zhang, Tianfang and Tang, Zijie and Shi, Cong and Wang, Yan and Chen, Yingying and Yuan, Bo},
  journal={arXiv preprint arXiv:2402.02739},
  year={2024}
}

@inproceedings{UFID,
  title={UFID: A Unified Framework for Black-box Input-level Backdoor Detection on Diffusion Models},
  author={Guan, Zihan and Hu, Mengxuan and Li, Sheng and Vullikanti, Anil Kumar},
  booktitle={Proceedings of the AAAI Conference on Artificial Intelligence},
  volume={39},
  number={26},
  pages={27312--27320},
  year={2025}
}

@inproceedings{Elijah,
  title={Elijah: Eliminating backdoors injected in diffusion models via distribution shift},
  author={An, Shengwei and Chou, Sheng-Yen and Zhang, Kaiyuan and Xu, Qiuling and Tao, Guanhong and Shen, Guangyu and Cheng, Siyuan and Ma, Shiqing and Chen, Pin-Yu and Ho, Tsung-Yi and others},
  booktitle={Proceedings of the AAAI Conference on Artificial Intelligence},
  volume={38},
  number={10},
  pages={10847--10855},
  year={2024}
}

@article{Diff-cleanse,
  title={Diff-cleanse: Identifying and mitigating backdoor attacks in diffusion models},
  author={Hao, Jiang and Jin, Xiao and Xiaoguang, Hu and Tianyou, Chen and Jiajia, Zhao},
  journal={arXiv preprint arXiv:2407.21316},
  year={2024}
}

@inproceedings{T2ishield,
  title={T2ishield: Defending against backdoors on text-to-image diffusion models},
  author={Wang, Zhongqi and Zhang, Jie and Shan, Shiguang and Chen, Xilin},
  booktitle={European Conference on Computer Vision},
  pages={107--124},
  year={2024},
  organization={Springer}
}

@article{CIFAR10,
  title={Learning multiple layers of features from tiny images},
  author={Krizhevsky, Alex and Hinton, Geoffrey and others},
  year={2009},
  publisher={Toronto, ON, Canada}
}

@inproceedings{D48,
  title={Deep learning face attributes in the wild},
  author={Liu, Ziwei and Luo, Ping and Wang, Xiaogang and Tang, Xiaoou},
  booktitle={Proceedings of the IEEE international conference on computer vision},
  pages={3730--3738},
  year={2015}
}

@inproceedings{Imagenet,
  title={Imagenet: A large-scale hierarchical image database},
  author={Deng, Jia and Dong, Wei and Socher, Richard and Li, Li-Jia and Li, Kai and Fei-Fei, Li},
  booktitle={2009 IEEE conference on computer vision and pattern recognition},
  pages={248--255},
  year={2009},
  organization={Ieee}
}

@inproceedings{Badencoder,
  title={Badencoder: Backdoor attacks to pre-trained encoders in self-supervised learning},
  author={Jia, Jinyuan and Liu, Yupei and Gong, Neil Zhenqiang},
  booktitle={2022 IEEE Symposium on Security and Privacy (SP)},
  pages={2043--2059},
  year={2022},
  organization={IEEE}
}

@inproceedings{B46,
  title={Backdoor attacks on self-supervised learning},
  author={Saha, Aniruddha and Tejankar, Ajinkya and Koohpayegani, Soroush Abbasi and Pirsiavash, Hamed},
  booktitle={Proceedings of the IEEE/CVF Conference on Computer Vision and Pattern Recognition},
  pages={13337--13346},
  year={2022}
}

@inproceedings{FID,
  author       = {Martin Heusel and
                  Hubert Ramsauer and
                  Thomas Unterthiner and
                  Bernhard Nessler and
                  Sepp Hochreiter},
  editor       = {Isabelle Guyon and
                  Ulrike von Luxburg and
                  Samy Bengio and
                  Hanna M. Wallach and
                  Rob Fergus and
                  S. V. N. Vishwanathan and
                  Roman Garnett},
  title        = {GANs Trained by a Two Time-Scale Update Rule Converge to a Local Nash
                  Equilibrium},
  booktitle    = {Advances in Neural Information Processing Systems 30: Annual Conference
                  on Neural Information Processing Systems 2017, December 4-9, 2017,
                  Long Beach, CA, {USA}},
  pages        = {6626--6637},
  year         = {2017},
}

@article{Dpm-solver,
  title={Dpm-solver: A fast ode solver for diffusion probabilistic model sampling in around 10 steps},
  author={Lu, Cheng and Zhou, Yuhao and Bao, Fan and Chen, Jianfei and Li, Chongxuan and Zhu, Jun},
  journal={Advances in neural information processing systems},
  volume={35},
  pages={5775--5787},
  year={2022}
}

@article{E53,
  title={Elucidating the design space of diffusion-based generative models},
  author={Karras, Tero and Aittala, Miika and Aila, Timo and Laine, Samuli},
  journal={Advances in neural information processing systems},
  volume={35},
  pages={26565--26577},
  year={2022}
}

@inproceedings{U-ViT,
  title={All are worth words: A vit backbone for diffusion models},
  author={Bao, Fan and Nie, Shen and Xue, Kaiwen and Cao, Yue and Li, Chongxuan and Su, Hang and Zhu, Jun},
  booktitle={Proceedings of the IEEE/CVF conference on computer vision and pattern recognition},
  pages={22669--22679},
  year={2023}
}

@inproceedings{Swin-unet,
  title={Swin-unet: Unet-like pure transformer for medical image segmentation},
  author={Cao, Hu and Wang, Yueyue and Chen, Joy and Jiang, Dongsheng and Zhang, Xiaopeng and Tian, Qi and Wang, Manning},
  booktitle={European conference on computer vision},
  pages={205--218},
  year={2022},
  organization={Springer}
}

@ARTICLE{SSIM,
  author={Zhou Wang and Bovik, A.C. and Sheikh, H.R. and Simoncelli, E.P.},
  journal={IEEE Transactions on Image Processing}, 
  title={Image quality assessment: from error visibility to structural similarity}, 
  year={2004},
  volume={13},
  number={4},
  pages={600-612},
  doi={10.1109/TIP.2003.819861}}

@inproceedings{Adam,
  author       = {Diederik P. Kingma and
                  Jimmy Ba},
  editor       = {Yoshua Bengio and
                  Yann LeCun},
  title        = {Adam: {A} Method for Stochastic Optimization},
  booktitle    = {3rd International Conference on Learning Representations, {ICLR} 2015,
                  San Diego, CA, USA, May 7-9, 2015, Conference Track Proceedings},
  year         = {2015},
  url          = {http://arxiv.org/abs/1412.6980},
  bibsource    = {dblp computer science bibliography, https://dblp.org}
}

@inproceedings{ViT,
  author       = {Alexey Dosovitskiy and
                  Lucas Beyer and
                  Alexander Kolesnikov and
                  Dirk Weissenborn and
                  Xiaohua Zhai and
                  Thomas Unterthiner and
                  Mostafa Dehghani and
                  Matthias Minderer and
                  Georg Heigold and
                  Sylvain Gelly and
                  Jakob Uszkoreit and
                  Neil Houlsby},
  title        = {An Image is Worth 16x16 Words: Transformers for Image Recognition
                  at Scale},
  booktitle    = {9th International Conference on Learning Representations, {ICLR} 2021,
                  Virtual Event, Austria, May 3-7, 2021},
  publisher    = {OpenReview.net},
  year         = {2021}
}

@article{FastDiff,
  title={Fastdiff: A fast conditional diffusion model for high-quality speech synthesis},
  author={Huang, Rongjie and Lam, Max WY and Wang, Jun and Su, Dan and Yu, Dong and Ren, Yi and Zhao, Zhou},
  journal={arXiv preprint arXiv:2204.09934},
  year={2022}
}

@article{HiddenSinger,
  title={HiddenSinger: High-quality singing voice synthesis via neural audio codec and latent diffusion models},
  author={Hwang, Ji-Sang and Lee, Sang-Hoon and Lee, Seong-Whan},
  journal={Neural Networks},
  volume={181},
  pages={106762},
  year={2025},
  publisher={Elsevier}
}

@inproceedings{PTID, author = {Chitwan Saharia and William Chan and Saurabh Saxena and Lala Li and Jay Whang and Emily L. Denton and Seyed Kamyar Seyed Ghasemipour and Raphael Gontijo Lopes and Burcu Karagol Ayan and Tim Salimans and Jonathan Ho and David J. Fleet and Mohammad Norouzi}, editor = {Sanmi Koyejo and S. Mohamed and A. Agarwal and Danielle Belgrave and K. Cho and A. Oh}, title = {Photorealistic Text-to-Image Diffusion Models with Deep Language Understanding}, booktitle = {Advances in Neural Information Processing Systems 35: Annual Conference on Neural Information Processing Systems 2022, NeurIPS 2022, New Orleans, LA, USA, November 28 - December 9, 2022}, year = {2022}, url = {http://papers.nips.cc/paper\_files/paper/2022/hash/ec795aeadae0b7d230fa35cbaf04c041-Abstract-Conference.html}, bibsource = {dblp computer science bibliography, https://dblp.org} }

@inproceedings{GLIDE, author = {Alexander Quinn Nichol and Prafulla Dhariwal and Aditya Ramesh and Pranav Shyam and Pamela Mishkin and Bob McGrew and Ilya Sutskever and Mark Chen}, editor = {Kamalika Chaudhuri and Stefanie Jegelka and Le Song and Csaba Szepesv{\'{a}}ri and Gang Niu and Sivan Sabato}, title = {{GLIDE:} Towards Photorealistic Image Generation and Editing with Text-Guided Diffusion Models}, booktitle = {International Conference on Machine Learning, {ICML} 2022, 17-23 July 2022, Baltimore, Maryland, {USA}}, series = {Proceedings of Machine Learning Research}, volume = {162}, pages = {16784--16804}, publisher = {{PMLR}}, year = {2022}, url = {https://proceedings.mlr.press/v162/nichol22a.html}, bibsource = {dblp computer science bibliography, https://dblp.org} }

@article{HTIG, author = {Aditya Ramesh and Prafulla Dhariwal and Alex Nichol and Casey Chu and Mark Chen}, title = {Hierarchical Text-Conditional Image Generation with {CLIP} Latents}, journal = {CoRR}, volume = {abs/2204.06125}, year = {2022}, url = {https://doi.org/10.48550/arXiv.2204.06125}, doi = {10.48550/ARXIV.2204.06125}, eprinttype = {arXiv}, eprint = {2204.06125}, bibsource = {dblp computer science bibliography, https://dblp.org} }

@inproceedings{SBG, author = {Yang Song and Jascha Sohl{-}Dickstein and Diederik P. Kingma and Abhishek Kumar and Stefano Ermon and Ben Poole}, title = {Score-Based Generative Modeling through Stochastic Differential Equations}, booktitle = {9th International Conference on Learning Representations, {ICLR} 2021, Virtual Event, Austria, May 3-7, 2021}, publisher = {OpenReview.net}, year = {2021}, url = {https://openreview.net/forum?id=PxTIG12RRHS}, bibsource = {dblp computer science bibliography, https://dblp.org} }

@inproceedings{GEG, author = {Yang Song and Stefano Ermon}, editor = {Hanna M. Wallach and Hugo Larochelle and Alina Beygelzimer and Florence d'Alch{\'{e}}{-}Buc and Emily B. Fox and Roman Garnett}, title = {Generative Modeling by Estimating Gradients of the Data Distribution}, booktitle = {Advances in Neural Information Processing Systems 32: Annual Conference on Neural Information Processing Systems 2019, NeurIPS 2019, December 8-14, 2019, Vancouver, BC, Canada}, pages = {11895--11907}, year = {2019}, url = {https://proceedings.neurips.cc/paper/2019/hash/3001ef257407d5a371a96dcd947c7d93-Abstract.html}, bibsource = {dblp computer science bibliography, https://dblp.org} }

@article{IPBA, author = {Jiayao Wang and Yang Song and Zhendong Zhao and Jiale Zhang and Qilin Wu and Junwu Zhu and Dongfang Zhao}, title = {{IPBA:} Imperceptible Perturbation Backdoor Attack in Federated Self-Supervised Learning}, journal = {CoRR}, volume = {abs/2508.08031}, year = {2025}, url = {https://doi.org/10.48550/arXiv.2508.08031}, doi = {10.48550/ARXIV.2508.08031}, eprinttype = {arXiv}, eprint = {2508.08031}, bibsource = {dblp computer science bibliography, https://dblp.org} }

@inproceedings{PDT, author = {Ren Wang and Gaoyuan Zhang and Sijia Liu and Pin{-}Yu Chen and Jinjun Xiong and Meng Wang}, editor = {Andrea Vedaldi and Horst Bischof and Thomas Brox and Jan{-}Michael Frahm}, title = {Practical Detection of Trojan Neural Networks: Data-Limited and Data-Free Cases}, booktitle = {Computer Vision - {ECCV} 2020 - 16th European Conference, Glasgow, UK, August 23-28, 2020, Proceedings, Part {XXIII}}, series = {Lecture Notes in Computer Science}, volume = {12368}, pages = {222--238}, publisher = {Springer}, year = {2020}, url = {https://doi.org/10.1007/978-3-030-58592-1\_14}, doi = {10.1007/978-3-030-58592-1\_14}, bibsource = {dblp computer science bibliography, https://dblp.org} }

@inproceedings{NC,
  title={Neural cleanse: Identifying and mitigating backdoor attacks in neural networks},
  author={Wang, Bolun and Yao, Yuanshun and Shan, Shawn and Li, Huiying and Viswanath, Bimal and Zheng, Haitao and Zhao, Ben Y},
  booktitle={2019 IEEE symposium on security and privacy (SP)},
  pages={707--723},
  year={2019},
  organization={IEEE}
}

@article{A31,
  title={Adversarial neuron pruning purifies backdoored deep models},
  author={Wu, Dongxian and Wang, Yisen},
  journal={Advances in Neural Information Processing Systems},
  volume={34},
  pages={16913--16925},
  year={2021}
}
}

\clearpage
\setcounter{page}{1}
\maketitlesupplementary
\setcounter{equation}{0}
\setcounter{figure}{0}
\setcounter{table}{0}
\setcounter{section}{0}
\setcounter{subsection}{0}

\section{Datasets}
\textbf{CIFAR-10~\cite{CIFAR10}:} It is a widely used benchmark dataset in machine learning, containing 60,000 color images (32$\times$32 pixels) divided into 10 classes. This dataset includes various common objects such as airplanes, cars, birds, cats, and dogs, making it well-suited for evaluating image classification models.\\
\textbf{CelebA-HQ~\cite{D48}:} It consists of 30,000 celebrity facial images with a high resolution of 1024$\times$1024 pixels. (In our work, we resized the images to 256$\times$256.) This dataset was created to improve the original CelebA~\cite{D48}, providing clearer and higher-resolution images, thereby enabling more accurate and robust model training in computer vision and generative tasks.\\
\textbf{CIFAR-100~\cite{CIFAR10}:} It is a widely used benchmark dataset in machine learning, containing 60,000 color images (32$\times$32 pixels) divided into 100 fine-grained classes. This dataset originates from the same source as CIFAR-10 but features a more detailed categorization: the 100 classes are organized into 20 superclasses, with each superclass containing 5 subclasses. The low resolution and multi-class nature of CIFAR-100 make it an ideal benchmark for evaluating fine-grained classification and generalization capabilities.\\
\textbf{ImageNet~\cite{Imagenet}:} It is a large-scale benchmark dataset in the field of computer vision, containing over 1.4 million annotated images (in the ILSVRC 2012 subset) spanning 1,000 categories. This dataset comprises natural images covering diverse classes such as animals, plants, objects, and scenes, with varying image resolutions. Due to its large scale and high category diversity, ImageNet has become the standard dataset for training and evaluating deep learning models, and is widely used in image classification and generation tasks. In our work, we resized the images to 256$\times$256 resolution for uniform processing.

\section{Detailed Explanation of Evaluation Metrics}
\textbf{Clean Accuracy (CA):} It measures whether the model's predictions or reconstructions on clean samples (without triggers) maintain the original semantics. A higher CA indicates that the attack causes less interference to the model's normal capabilities.\\
\textbf{Backdoor Accuracy (BA):} It measures the accuracy of the model producing the expected target semantics on samples containing triggers. A higher BA indicates more stable and reliable behavior when the trigger is activated.\\
\textbf{Attack Success Rate (ASR):} It measures the proportion of times the model is guided to the attacker-specified output when the trigger is present. ASR can be viewed as a specific manifestation of BA in certain task definitions. An effective backdoor attack should maximize ASR/BA while maintaining high CA.\\
\textbf{Fréchet Inception Distance (FID)~\cite{FID}:} It measures the distribution difference between model outputs and clean data in generative tasks. A lower value indicates that the generated clean samples are closer to the real data distribution.\\
\textbf{Mean Squared Error (MSE):} For samples containing triggers, MSE measures the average squared difference between the model output and the target image. A lower value indicates higher precision in pixel-level reconstruction for the attack.\\
\textbf{Structural Similarity Index Measure (SSIM)~\cite{SSIM}:} It evaluates the consistency between the output of triggered samples and the target image in terms of luminance, contrast, and structure. SSIM ranges from [0,1], with values closer to 1 indicating better structural similarity to the target.

\section{Implementation Details and Parameter Settings of BadRSSD}
\textbf{DiT Architecture:} RSSD and BadRSSD: Both are based on the standard DiT-Large (DiT-L/2-SSL), which is a self-supervised learning variant of DiT-L~\cite{S19}. DiT-L/2-SSL uses a half-depth ViT-L architecture: 12-layer encoder and 12-layer decoder, with the same total depth as ViT-L (24 blocks)~\cite{ViT}, and an MLP ratio of 1/4 of the original dimension. BadRSSD is evaluated on four architectures: DiT-L/2~\cite{S19}, DiT-XL/2~\cite{S19}, U-ViT~\cite{U-ViT}, and Swin-UNet~\cite{Swin-unet}, to verify method generalization.\\
\textbf{Training:} RSSD: Adam optimizer, learning rate 1e-4, batch size 2048, 400 training epochs, weight decay 0.0, linear warmup for 100 epochs followed by half-cycle cosine decay, dataset ImageNet-1K (256×256). BadRSSD: Adam optimizer, learning rate 1e-4, batch size 32, 100 training epochs (partially 400 epochs for long-term stability evaluation), weight decay 0.0, warmup 10 epochs, datasets CIFAR10 (32×32), CIFAR100 (32×32), CelebA-HQ (256×256), ImageNet (256×256). Backdoor attack configuration: trigger size 16×16 (or 4×4) pixels (bottom-right corner), poisoning rate 5\%, alignment strategy direct alignment.\\
\textbf{Loss Functions and Weight Parameters:} RSSD: Collaborative dual loss $L_{\text{RSSD}} = L_{\text{SSL}} + 0.5 \cdot L_{\text{disp}}$, where $L_{\text{SSL}}$ computes semantic alignment in PCA space, $L_{\text{disp}}$ extracts features from DiT intermediate layers (layer 3) to constrain representation dispersion, temperature parameter $\tau=1$. BadRSSD: Conditional triple loss function, clean samples use the same loss as RSSD ($L_{\text{SSL}} + 0.5 \cdot L_{\text{disp}}$), poisoned samples use triple loss ($2.0 \cdot L_{\text{PCA\_TR}} + 1.5 \cdot L_{\text{img\_rec}} + 0.5 \cdot L_{\text{disp}}$), where $L_{\text{PCA\_TR}}$ implements PCA space alignment and trajectory consistency, trajectory consistency weight $\beta=0.5$, $L_{\text{img\_rec}}$ ensures pixel-level attack control.\\
\textbf{Sampling:} RSSD and BadRSSD: Both use DPM-solver sampling method (20 steps), uniform time-step sampler, linear noise schedule ($\alpha = 0.1$). BadRSSD processes triggered inputs through PCA alignment during inference to implement backdoor attacks.\\
\textbf{Evaluation Metrics:} RSSD: FID-50k evaluates generation quality, linear probing accuracy evaluates representation quality. BadRSSD: FID measures utility (performance on clean inputs without triggers), MSE quantifies specificity (accuracy of generating target images from triggered inputs), ASR evaluates attack success rate, SSIM evaluates attack stealthiness (structural similarity between poisoned and clean samples).\\
\textbf{Datasets:} RSSD: ImageNet-1K (256×256), minimal data augmentation. BadRSSD: CIFAR10 (32×32), CIFAR100 (32×32), CelebA-HQ (256×256), ImageNet (256×256), covering low and high resolutions to verify cross-resolution stability.\\
\textbf{PCA Configuration:} RSSD and BadRSSD: Patch size 16×16 (or 4×4), PCA latent dimension 48, total PCA dimension 768, noise schedule $\sigma_t = \alpha \cdot t/T$ ($\alpha = 0.1$). BadRSSD adds PCA space backdoor alignment operation ($z_0^{a} = z_0^{P} + \Delta z_{t}$), PCA space backdoor alignment strength 1.0, to achieve semantic alignment.\\
\textbf{Experimental Conclusions:} RSSD and BadRSSD are consistent in DiT architecture, sampling methods, and PCA configuration, ensuring BadRSSD can fully utilize RSSD's representation learning capability. Through conditional triple loss function, PCA space backdoor alignment, and multi-dataset evaluation, BadRSSD achieves efficient backdoor attacks while maintaining RSSD's representation learning capability and generation quality.

\section{Implementation details of the four architectures: DiT-L/2, DiT-XL/2, U-ViT, and Swin-UNet}
The core innovations of the RSSD and BadRSSD frameworks (PCA space noise injection, conditional loss function, Dispersive Loss) are architecture-agnostic and can be adapted to different Transformer diffusion models. Alth-ough the base implementation of RSSD and BadRSSD is built on the DiT-L/2-SSL architecture (half-depth ViT-L, 12-layer encoder + 12-layer decoder, see implementation details in the main text), to verify the method's universality and generalization capability, we adapt the framework to the following standard architectures for evaluation. During adaptation, we primarily adjust the position of feature extraction layers (for Dispersive Loss calculation) and the settings of PCA block sizes, while keeping the core algorithms (PCA space noise injection, conditional loss function design) unchanged.
\subsection{DiT-L/2 Architecture}
DiT-L/2~\cite{S19} is a medium-scale single-scale full Transformer diffusion model using standard DiT architecture design (24 consecutive Transformer blocks, no encoder/decoder separation). Input images are converted to token sequences through small patch embedding (/2 indicates patch size of 2), processed by 24 Transformer blocks, each containing multi-head self-attention and MLP. Timestep t is modulated through AdaLN (Adaptive Layer Normalization) for channel-wise conditioning in each block. The model employs standard diffusion training objective (predicting clean data x0).

RSSD/BadRSSD Adaptation: Adapting RSSD framework core components to standard DiT-L/2 architecture: PCA Space Noise Injection: Through block-wise PCA processing (patch size=16), injecting noise in PCA low-dimensional space, then inverse PCA mapping back to image space, compatible with standard DiT patch embedding mechanism. Conditional Loss Function: Clean samples use $L_{\text{SSL}} + 0.5 \cdot L_{\text{disp}}$, poisoned samples use $2.0 \cdot L_{\text{PCA\_trajectory}} + 1.5 \cdot L_{\text{img\_recon}} + 0.5 \cdot L_{\text{disp}}$, independent of specific architectural structure. Dispersive Loss: Extract intermediate features from the 12th Transformer block (middle position of 24 layers), compute dispersion loss after global average pooling, ensuring feature extraction position aligns with base architecture (encoder-decoder boundary in DiT-L/2-SSL). This architecture balances image generation quality and computational efficiency, suitable as baseline model for multi-architecture validation, verifying RSSD framework adaptability on standard DiT architecture.

\subsection{DiT-XL/2 Architecture}
DiT-XL/2~\cite{S19} is the large-capacity variant in DiT series, using standard DiT architecture design (28 consecutive Transformer blocks, no encoder/decoder separation), with same structure as DiT-L/2 but larger scale (hidden\_size=1152, depth=28, num\_heads=16). Larger model capacity enables lower FID and MSE under same training epochs, demonstrating stronger image generation capability and reconstruction accuracy. Training employs gradient accumulation, mixed precision and zero-init initialization strategies to ensure stability.

RSSD/BadRSSD Adaptation: RSSD framework adaptation method same as DiT-L/2, but optimized for larger-scale model: PCA Space Noise Injection: Using same block-wise PCA processing mechanism, adapted to DiT-XL/2 patch embedding. Conditional Loss Function: Using same loss function design, leveraging larger model capacity to enhance attack effectiveness. Dispersive Loss: Extract intermediate features from the 14th Transformer block (middle position of 28 layers), adopting denser feature extraction strategy (e.g., every 4 layers), fully utilizing deep representations, verifying BadRSSD effectiveness on large-scale models. This architecture suits scenarios pursuing optimal performance, verifying BadRSSD effectiveness on large-scale models and RSSD framework scalability.

\subsection{U-ViT Architecture}
U-ViT~\cite{U-ViT} employs encoder-bottleneck-decoder U-Net topology, transforming ViT into multi-scale hierarchical architecture. Encoder performs gradual downsampling thr-ough Patch Merging (resolution decreases, channel number increases), stacking Transformer blocks at each level to extract multi-scale features; bottleneck layer consists of multiple Transformer blocks carrying global context information; decoder performs gradual upsampling through PixelShuffle or transposed convolution, fusing with corresponding encoder level features via skip connections to achieve detail restoration. Timestep conditioning is injected through AdaLN in multi-scale Transformer blocks.

RSSD/BadRSSD Adaptation: U-ViT's multi-scale structure provides rich adaptation points for RSSD framework: PCA Space Noise Injection: Perform block-wise PCA processing at encoder input stage, adapting to U-ViT multi-scale patch embedding mechanism. Since U-ViT uses different patch sizes at different scales, PCA block size needs adaptive adjustment according to current scale. Conditional Loss Function: Using same conditional loss function design, leveraging U-ViT multi-scale features to enhance attack effectiveness. Dispersive Loss: Extract multi-scale intermediate features from encoder levels, bottleneck layer and decoder levels, providing rich intermediate representations for Dispersive Loss. Feature extraction positions select middle encoder levels (e.g., encoder level 3) and bottleneck layer, ensuring consistency with base architecture feature extraction strategy. This architecture naturally outputs multi-scale features (encoder layers + bottleneck + decoder layers), combining Transformer's global modeling capability with U-Net's multi-scale fusion advantage, demonstrating good performance and convergence speed in multi-architecture validation, verifying RSSD framework adaptability on multi-scale architectures.

\subsection{Swin-UNet Architecture}
Swin-UNet~\cite{Swin-unet} is based on Swin Transformer hierarchical pyramid structure, employing window attention mechanism (Window-based Multi-head Self-Attention, Win-MSA) and shifted window attention (Shifted Win-MSA) alternately applied, achieving local window self-attention computation and cross-window information interaction. Encoder establishes hierarchical structure through Patch Merging (C→2C→4C→8C), decoder performs gradual upsampling through Patch Expanding and fuses with corresponding encoder layer skip connections. This architecture doesn't rely on explicit position encoding, but implicitly encodes spatial information through relative position bias. Timestep conditioning is injected through AdaLN in Swin Block, consistent with DiT series. Window attention mechanism makes computational complexity approximately linear to image resolution, significantly saving memory and computation time in high-resolution scenarios.

RSSD/BadRSSD Adaptation: Swin-UNet's hierarchical structure provides efficient adaptation scheme for RSSD framework: PCA Space Noise Injection: Perform block-wise PCA processing at encoder input stage, adapting to Swin-UNet hierarchical patch embedding mechanism. Since Swin-UNet uses different window sizes and channel numbers at different levels, PCA block size needs adaptive adjustment according to current level. Conditional Loss Function: Using same conditional loss function design, leveraging Swin-UNet hierarchical features to enhance attack effectiveness. Dispersive Loss: Extract hierarchical intermediate features from encoder levels (C, 2C, 4C, 8C), selecting middle encoder levels (e.g., 2C or 4C level) for feature extraction, ensuring consistency with base architecture feature extraction strategy. Swin-UNet hierarchical features naturally adapt to multi-scale Dispersive Loss computation. This architecture verifies BadRSSD effectiveness on efficient Transformer architectures, window attention mechanism efficiency enables RSSD framework to maintain computational efficiency in high-resolution scenarios, verifying method practicality and scalability.

\subsection{Summary}
Evaluation experiments on the above four architectures demonstrate that the core components of RSSD and BadRSSD frameworks (PCA space noise injection, conditional loss function, Dispersive Loss) possess good architecture independence and universality, and can be adapted to different types of Transformer diffusion model architectures. During adaptation, mainly adjusting feature extraction layer positions and PCA block size settings, while keeping core algorithms unchanged, this proves the rationality of RSSD framework design and method generalizability.

\section{Robustness Experiment Explanation and Evaluation Metric Calculation Details}

\subsection{DisDet}
DisDet~\cite{Disdet} (Distribution-based Detection) uses distribution differences: detecting marginal statistical differences between clean and poisoned samples. Premise: backdoor attacks cause detectable distribution differences. Process: Distribution difference calculation (compute marginal statistical features for clean/poisoned samples to get PDD, larger PDD indicates more significant differences); Detection model training (train binary classifier using PDD features to distinguish samples); Backdoor sample handling (remove triggers from or reject detected poisoned samples).

Evaluation metrics:
\begin{itemize}
\item PDD(Clean) and PDD(Poison): Distribution difference scores for clean/poisoned samples, computed from marginal statistical features (mean, variance, higher-order moments, etc.), difference $\Delta \text{PDD} = \text{PDD(Poison)} - \text{PDD(Clean)}$, larger $\Delta \text{PDD}$ indicates more significant differences
\item AUROC: Area under ROC curve, computed by varying detection threshold to calculate TPR and FPR at different thresholds then plotting ROC curve, 1.0 perfect, 0.5 random
\item TPR@1\%FPR: TPR at fixed FPR=1\%, obtained by adjusting threshold to achieve FPR=0.01 then reading corresponding TPR, higher values indicate better detection capability at low false positive rates
\item Detection Pass Rate: Percentage of poisoned samples passing detection (escape rate), $\text{Detection Pass Rate} = (1 - \text{TPR}) \times 100\%$, lower values indicate better defense effectiveness
\item $ASR_{before}$ and $ASR_{after}$: Attack success rate before/after defense, $\text{ASR} = \frac{1}{N} \sum_{i=1}^N \mathbb{I} \left[ \text{SSIM} \left( f \left( x_i^{\text{poisoned}} \right), y_{\text{target}} \right) > \tau_{\text{ASR}} \right]$, defense effectiveness measured by $\Delta \text{ASR} = \text{ASR}_{before} - \text{ASR}_{after}$
\item $\Delta$FID: Generation quality change after defense, $\Delta \text{FID} = \text{FID}_{after} - \text{FID}_{before}$, where $\text{FID} = \|\mu_{real} - \mu_{gen}\|_2^2 + \text{Tr}(\Sigma_{real} + \Sigma_{gen} - 2(\Sigma_{real}\Sigma_{gen})^{1/2})$
\end{itemize}

Results: Effective against BadDiffusion and TrojDiff, $\Delta \text{PDD} \approx 0.45-0.44$, AUROC 0.92-0.95, TPR@1\%FPR 85.46-87.12\%, Detection Pass Rate 26.23-32.17\%, ASR reduced from 71.02-73.18\% to 8.26-9.45\% ($\Delta \text{ASR} \approx 62-64\%$). Ineffective against BadRSSD: $\Delta \text{PDD} = 0.12$, AUROC 0.58 (near random), TPR@1\%FPR 8.72\%, Detection Pass Rate 85.68\%, ASR reduced from 94.67\% to only 92.57\% ($\Delta \text{ASR} = 2.10\%$), $\Delta \text{FID} = 4.63$ (positive). Failure reasons: BadRSSD maintains marginal distribution stability through $L_{disp}$, clean/poisoned samples have almost identical statistical features; backdoor effects mainly in encoded PCA subspace and later timesteps, conditional differences at low poisoning rates get "averaged" into marginal statistics, resulting in small PDD separation; through PCA space alignment and dispersion loss, backdoor effects don't produce significant differences at marginal statistical level.

\subsection{Elijah}
Elijah~\cite{Elijah} is a backdoor defense method based on trigger inversion and neuron pruning. Process: Trigger inversion ($r^* = \arg\min_{r} \sum_{i=1}^{N} \mathcal{L}(f(x_i + r), y_{target})$), Neuron localization (analyze neuron responses to triggers, locate relevant neuron clusters), Neuron pruning (prune relevant neuron clusters to remove backdoor).

Evaluation metrics:
\begin{itemize}
\item Poison Rate: $\text{Poison Rate} = \frac{N_{poisoned}}{N_{total}} \times 100\%$, table uses six poisoning rates: 5\%, 10\%, 20\%, 30\%, 40\%, 50\%
\item Detected(\%): $\text{Detected}(\%) = \frac{N_{detected}}{N_{total}} \times 100\%$, higher detection rate indicates more accurate backdoor identification
\item $ASR_{before}$ and $ASR_{after}$: Attack success rate before/after defense, $\text{ASR} = \frac{1}{N} \sum_{i=1}^N \mathbb{I} \left[ \text{SSIM} \left( f \left( x_i^{\text{poisoned}} \right), y_{\text{target}} \right) > \tau_{\text{ASR}} \right]$, defense effectiveness measured by $\Delta \text{ASR} = \text{ASR}_{before} - \text{ASR}_{after}$
\item $\Delta$SSIM: Change in structural similarity between generated and target images after defense, $\Delta \text{SSIM} = \text{SSIM}_{after} - \text{SSIM}_{before}$, where $\text{SSIM}(x, y) = \frac{(2\mu_x\mu_y + c_1)(2\sigma_{xy} + c_2)}{(\mu_x^2 + \mu_y^2 + c_1)(\sigma_x^2 + \sigma_y^2 + c_2)}$, $\Delta \text{SSIM} < 0$ indicates backdoor function removed, $\Delta \text{SSIM} \approx 0$ indicates backdoor function not removed
\item $\Delta$FID: Generation quality change after defense, $\Delta \text{FID} = \text{FID}_{after} - \text{FID}_{before}$, where $\text{FID} = \|\mu_{real} - \mu_{gen}\|_2^2 + \text{Tr}(\Sigma_{real} + \Sigma_{gen} - 2(\Sigma_{real}\Sigma_{gen})^{1/2})$
\end{itemize}

Results: Effective against BadDiffusion, detection rate increased from 82.06\% to 100\%, ASR reduced from 71.02-92.65\% to 4.02-6.07\% ($\Delta \text{ASR} \approx 65-87\%$), $\Delta \text{SSIM} \approx -0.8 \sim -1.0$, $\Delta \text{FID}$ small and changes insignificant. Ineffective against BadRSSD: very low detection rate (5.12-18.08\%), ASR reduced from 94.67-99.99\% to only 92.57-98.65\% ($\Delta \text{ASR} \approx 1-3\%$), $\Delta \text{SSIM} \approx 0$, $\Delta \text{FID}$ positive and increases with poisoning rate (-0.89 $\sim$ 8.5). Elijah fails against BadRSSD because: effective trigger direction resembles small, unstructured global perturbations difficult to invert into stable generalizable triggers; carrying pathways dispersed across layers and time, no concentrated neuron clusters for pruning; backdoor effects non-local and dispersed, cannot be removed by pruning specific neurons. Thus, BadRSSD's PCA space semantic alignment and full-time trajectory drive disperses backdoor effects across representation space and time dimensions, making Elijah's trigger inversion and neuron pruning ineffective for detection and removal, demonstrating the stealthiness and threat of representation-layer backdoor attacks.

\subsection{TERD}
TERD~\cite{TERD} (Trigger-based Reverse Engineering Defense) is based on trigger inversion: reverse-engineering generalizable structured triggers from model outputs for detection and mitigation. The premise is that backdoors use fixed structured triggers (e.g., patches) detectable in pixel domain. Process: Trigger inversion $r^* = \arg\min_{r} \mathbb{E}_{x \sim \mathcal{D}} \left[ \mathcal{L}(f(x + r), y_{target}) \right] + \lambda \cdot \mathcal{R}(r)$ ($\lambda$ is parameter $\eta$ in table); Trigger detection: use $r^*$ to determine if input contains backdoor ($\mathcal{L}(f(x + r^*), y_{target}) < \tau$); Trigger removal: perform $x_{cleaned} = x - r^*$ on detected samples.

Evaluation metrics:
\begin{itemize}
\item $ASR_{before}$ and $ASR_{after}$: Attack success rate before/after defense, $\text{ASR} = \frac{1}{N} \sum_{i=1}^N \mathbb{I} \left[ \text{SSIM} \left( f \left( x_i^{\text{poisoned}} \right), y_{\text{target}} \right) > \tau_{\text{ASR}} \right]$, $\Delta \text{ASR} = \text{ASR}_{before} - \text{ASR}_{after}$ measures defense effectiveness
\item $\Delta$FID: Generation quality change after defense, $\Delta \text{FID} = \text{FID}_{after} - \text{FID}_{before}$, where $\text{FID} = \|\mu_{real} - \mu_{gen}\|_2^2 + \text{Tr}(\Sigma_{real} + \Sigma_{gen} - 2(\Sigma_{real}\Sigma_{gen})^{1/2})$
\item TPR: True positive rate, $\text{TPR} = \frac{\text{TP}}{\text{TP} + \text{FN}}$
\item TNR: True negative rate, $\text{TNR} = \frac{\text{TN}}{\text{TN} + \text{FP}}$
\item $\|r-r_0\|_2$: L2 distance between inverted and real triggers, $\|r - r_0\|_2 = \sqrt{\sum_{i,j,k} (r_{i,j,k} - r_{0,i,j,k})^2}$
\end{itemize}

Results: Effective against BadDiffusion: ASR reduced from 71.02-68.19\% to 8.12-7.65\% ($\Delta \text{ASR} \approx 60-63\%$), TPR 76.56-85.12\%, $\|r-r_0\|_2$ 0.18-0.26. Ineffective against BadRSSD: ASR reduced from 94.67-92.51\% to only 92.35-91.76\% ($\Delta \text{ASR} \approx 0.75-2.32\%$), TPR 4.28-6.74\%, $\|r-r_0\|_2$ 0.75-0.82, $\Delta \text{FID}$ 4.52-8.16. TNR 96-97\% for both, indicating normal identification of clean samples; problem lies in difficulty detecting poisoned samples. TERD fails against BadRSSD due to assumption-mechanism mismatch: TERD assumes fixed structured triggers, while BadRSSD is driven by PCA semantic alignment and full-time trajectory with non-local, dispersed activation directions; TERD assumes pixel-domain detectability, while BadRSSD effects mainly in encoded PCA subspace and later timesteps, causing misalignment between pixel-domain inversion and real causes; TERD assumes localized triggers, while BadRSSD effective triggering resembles small, unstructured global perturbations difficult to invert into stable generalizable triggers.

\end{document}